\pdfoutput=1
\documentclass[12pt, reqno]{article}

\usepackage{amsmath}
\usepackage{cite}
\usepackage{hyperref}
\usepackage{epsfig}
\usepackage{amssymb}
\usepackage{color}
\usepackage{mathrsfs}

\newcommand{\be}{\begin{eqnarray}}
\newcommand{\ee}{\end{eqnarray}}

\renewcommand{\bar}[1]{\overline{#1}}

\newcommand{\tm}{\mathrm{\rule[2.4pt]{6pt}{0.65pt}}}

\renewcommand{\tilde}[1]{\widetilde{#1}}
\newcommand{\smallminus}{{\rm\rule[2.4pt]{6pt}{0.65pt}}}
\newcommand{\smallplus}{\hspace{0.5pt}\text{{\small+}}\hspace{-0.5pt}}
\definecolor{downstairs}{rgb}{0.09, 0.1328, 0.7888}
\definecolor{upstairs}{rgb}{0.7598,0.1259,0.259}
\newcommand{\softdown}[1]{_{\raisebox{-1.95pt}{{\footnotesize{\color{downstairs}#1}}}}}
\newcommand{\softup}[1]{^{\raisebox{0.05pt}{{\footnotesize{\color{upstairs}#1}}}}}
\newcommand{\soft}[2]{\softup{#1}\softdown{#2}\!}
\newcommand{\inversesoft}[2]{\underset{#1}{\phantom{{}^{#2}}{\mathcal{S}^{#2}}}}
\newcommand{\vf}{\boldsymbol{f}}
\newcommand{\vS}{\boldsymbol{S}}

\makeindex
\begin{document}

\baselineskip=18pt

\setcounter{footnote}{0}
\setcounter{figure}{0}
\setcounter{table}{0}

\begin{titlepage}

\begin{center}

{\Large \bf Unification of Residues and Grassmannian Dualities}
\vspace{0.1cm}

{\bf N. Arkani-Hamed$^a$, J. Bourjaily$^{a,c}$, F. Cachazo$^{a,b}$
and J. Trnka$^{a,c}$}

\vspace{.1cm}

{\it $^{a}$ School of Natural Sciences, Institute for Advanced Study, Princeton, NJ 08540, USA}

{\it $^{b}$ Perimeter Institute for Theoretical Physics, Waterloo, Ontario N2J W29, CA}

{\it $^{c}$ Department of Physics, Princeton University, Princeton, NJ 08544, USA}

\end{center}

\begin{abstract}

The conjectured duality relating all-loop leading singularities of $n$-particle
N$^{k-2}$MHV scattering amplitudes in ${\cal N}=4$ SYM to a simple
contour integral over the Grassmannian $G(k,n)$ makes all the
symmetries of the theory manifest. Every residue is individually
Yangian invariant, but does not have a local space-time
interpretation---only a special sum over residues gives physical
amplitudes. In this paper  we show that the sum over residues giving
tree amplitudes can be unified into a single algebraic variety, which
we explicitly construct for all
NMHV and N$^2$MHV amplitudes. Remarkably, this allows the contour integral to have
a ``particle interpretation" in
the Grassmannian, where higher-point amplitudes can be constructed from lower-point ones by
adding one particle at a time, with soft limits manifest. 
We move on to show that the connected prescription for tree
amplitudes in Witten's twistor string theory also admits a
Grassmannian particle interpretation, where the integral over the
Grassmannian localizes over the Veronese map from  $G(2,n) \to
G(k,n)$. These apparently very different
theories are related by a natural deformation with a parameter $t$ that smoothly interpolates between them. For NMHV amplitudes, we use a simple
residue theorem to prove $t$-independence of the result, thus
establishing a novel kind of duality between these
theories.

\end{abstract}

\bigskip
\bigskip

\end{titlepage}

\section{Scattering Amplitudes and the Grassmannian}\setcounter{equation}{0}

A new duality has recently been conjectured \cite{ArkaniHamed:2009dn} between leading singularities of color-stripped $n$-particle N$^{k-2}$MHV amplitudes in ${\cal N}=4$ SYM and a simple contour integral of the form
\be\label{eq1}
{\cal L}_{n,k}({\cal W}_a) = \frac{1}{{\rm vol (GL}(k))}\int \frac{d^{k\times n}C_{\alpha a}}{(1\,2\cdots k)(2\,3\cdots k\smallplus1)\cdots (n\,1\cdots k\smallminus1)} \prod_{\alpha =1}^k \delta^{4|4}(\sum_{a=1}^nC_{\alpha a}{\cal W}_a),
\ee
where the $\mathcal{W}_a$ in the (ordinary) dual twistor space and carry all the information about the external particles. The integral is over $k\times n$ matrices $C_{\alpha\,a}$ modulo a GL$(k)$-action on the right. This space is also known as the Grassmannian
$G(k,n)$---the space of configurations of $k$-planes in $\mathbb{C}^n$. The rows in the matrix $C_{\alpha\,a}$
define $k$ $n$-vectors which together span a $k$-plane that contains the
origin. Since GL$(k)$-transformations simply reflect a change of basis for the $k$-plane, the action of GL$(k)$ must be modded-out. The formulation in (\ref{eq1}) makes manifest that any object computed from ${\cal
L}_{n,k}$ is superconformal invariant. 

Fourier-transforming from dual twistors to ordinary momentum-space, one finds that
\begin{equation}
\label{eq:mome}
\begin{split}{\cal L}_{n,k}&=\displaystyle \frac{1}{{\rm vol(GL}(k))}\int \frac{d^{k\times n}C}{(1\,2\cdots k)(2\,3\cdots k\smallplus1)\cdots (n\,1\cdots k\smallminus1)} \\ & \qquad\qquad\qquad\qquad\times  \prod_{\alpha=1}^k \delta^4(C_{\alpha a}\tilde\eta_a) \delta^2(C_{\alpha a}\tilde\lambda_a) \int d^2\rho_\alpha \delta^2(\rho_\beta C_{\beta a}-\lambda_a)\,.
\end{split}
\end{equation}
Gauge-fixing the GL$(k)$ redundancy in such a way that $k$
columns of the matrix $C_{\alpha\,a}$ make up the unit $k\times k$ matrix
takes (\ref{eq:mome}) into the link representation of
\cite{ArkaniHamed:2009si}. This gauge-fixing makes parity manifest by making it equivalent to the obvious geometric statement that
$G(k,n)$ is isomorphic to $G(n-k,n)$. The $\delta$-functions in
(\ref{eq:mome}) restrict the integration to $k$-planes that contain
the $\lambda$-plane and are orthogonal to the $\tilde\lambda$-plane.
Using a different gauge-fixing, one can make the first two rows of
the $C$-matrix be identical to the two $n$-vectors defining the
$\lambda$-plane. A simple linear algebra argument together with a
further gauge fixing that leaves a GL$(k-2)$ subgroup of GL$(k)$
unfixed reduces the integral to one over $(k-2)$-planes in
$\mathbb{C}^n$, i.e.\ , over $G(k-2,n)$ \cite{ArkaniHamed:2009vw}. The
resulting form, in terms of a $(k-2)\times n$ matrix $D$ is given by
\cite{ArkaniHamed:2009vw,Mason:2009qx},
\be\label{second}
\hspace{-0.03cm}{\cal L}_{n,k} = {\cal A}_{\rm MHV}\frac{1}{{\rm vol(GL}(k-2))}\int \frac{d^{(k-2)\times n}D}{(1\,2\cdots k\smallminus2)(2\,3\cdots k\smallminus1)\cdots (n\,1\cdots k\smallminus3)} \prod_{\hat{\alpha}=1}^{k-2} \delta^{4|4}(D_{\hat{\alpha} a}{\cal Z}_a),\quad
\ee
where ${\cal A}_{\rm MHV}$ is the tree-level MHV superamplitude
which contains  the momentum-conserving $\delta$-function and its
superpartner. The remaining integral is now defined in terms of what
are called momentum-supertwistors ${\cal Z}_a$. These are the
objects introduced by Hodges \cite{Hodges:2009hk} in order to make
{\it dual}-superconformal invariance \cite{Drummond:2006rz,
Alday:2007hr, Elvang:2009ya, Brandhuber:2008pf} manifest.

After all $\delta$-functions in (\ref{eq:mome}) are used, ${\cal
L}_{n,k}$ becomes a contour integral in \mbox{$(k-2)(n-k-2)$}
variables. As usual with contour integrals, there is really no
integral at all and we are interested in the residues. Each of these
residues is simultaneously superconformal and dual-superconformal
invariant, and is thus invariant under the full Yangian symmetry of the theory \cite{Drummond:2008vq,Drummond:2009fd}.
Higher-dimensional analogues of Cauchy's residue theorem encode highly non-trivial
relations between these invariants. The residues give a basis for
the leading singularities of all loop amplitudes. Evidence for this fact for up to two-loops was given in \cite{ArkaniHamed:2009dn}, and evidence to all orders has been recently given by \cite{Bullimore:2009cb,Kaplan:2009mh}.  Tree-level
amplitudes are known to be expressible as sums over one-loop
leading singularities---via the  BCFW recursion relations
\cite{Britto:2004ap, Britto:2005fq} \mbox{(see also, e.g,.
\cite{Drummond:2008cr})}---and therefore they become sums of residues
of ${\cal L}_{n,k}$. This can be expressed by providing a contour of
integration for ${\cal L}_{n,k}$ which we denote $\Gamma^{{\cal L}}_{n,k}$. Note that this contour is not uniquely defined, since residue theorems
can be used to express the same sum in many different forms. We will nonetheless loosely refer to this equivalence class of contours as ``the" contour.

The contour $\Gamma_{n,k}^{{\cal L}}$ must have a remarkable
property. While the residues are all Yangian invariant, they do not
individually have a local space-time interpretation; for
instance, they are riddled with non-local poles. The non-local poles
magically cancel in the sum over residues of $\Gamma_{n,k}$. In our
previous paper \cite{ArkaniHamed:2009sx}, we showed that a natural
contour deformation ``blows up residues" into a sum over local and
non-local terms, making the local spacetime description as manifest
as possible by connecting to the light-cone gauge Lagrangian via the
CSW/Risager \cite{Risager:2005vk, Risager:2008yz, Cachazo:2004zb,
Cachazo:2004kj, Cachazo:2004by} rules. In this paper we discuss a
natural counterpart to this operation: instead of ``blowing up"
residues, we will see that there is a natural way of {\it unifying}
them into a single algebraic variety. This will expose something
perhaps even more surprising than the emergence of local space-time
physics: we will see that the contour $\Gamma_{n,k}^{{\cal L}}$
can be thought of as localizing the integral over $G(k,n)$ to a
sub-manifold with a ``particle interpretation" {\it in the
Grassmannian}. This allows us to construct higher-point tree
amplitudes by simply ``adding one particle at a time'' to lower-point ones,
with soft limits manifest. Furthermore, this unified form of the amplitude 
is intimately connected to CSW localization in 
twistor space, and---as we will see for N$^2$MHV---is generally distinct from any contour derived using BCFW.

\pagebreak
Having discovered the possibility of a particle interpretation in
the Grassmannian, it is natural to ask whether there is a
formulation that makes such an interpretation manifest while also keeping manifest cyclic invariance (which would not ordinarily be completely explicit in a picture which ``adds one particle at a time''). This
motivates us to start anew, keeping only the Grassmannian kinematics
encoded in the $\delta$-function factor $\delta^{4|4}(C_{\alpha a}
{\cal W}_a)$. A simple counting argument leads us to an extremely
natural way of implementing the Grassmannian particle interpretation:
by integrating over a sub-manifold in the Grassmannian associated
with the ``Veronese map" from $G(2,n) \to G(k,n)$. The resulting
object can be easily recognized as the connected prescription
\cite{Roiban:2004yf} for Witten's twistor string theory
\cite{Witten:2003nn} (see also \cite{Berkovits:2004hg, Dolan:2009wf,
Spradlin:2009qr, Roiban:2004ka, Witten:2004cp, Vergu:2006np,
Gukov:2004ei, Dixon:2005cf, Mason:2005zm, Berkovits:2004jj,
Dolan:2008gc, Bedford:2007kr, Mason:2007zv, Dolan:2007vv}; for
a review, see \cite{Cachazo:2005ga}); indeed this discussion can be thought
of as a physical motivation for and derivation of this theory from
the Grassmannian viewpoint.

Cast as integrals over the Grassmannian, the integrand corresponding to our first discovery of the particle interpretation---motivated by realizing the contour $\Gamma^{{\cal L}}_{n,k}$ as a single algebraic variety---will not be the same as the second form, leading to the connected prescription for twistor string theory. In the simplest examples, one can use the global residue theorem \mbox{(see e.g. \cite{Griffiths:1978a})} to show that while the integrands are different, the contour integrals agree \mbox{(see e.g. \cite{Nandan:2009cc})}. However, this way of establishing the equality requires some gymnastics; a significant insight into why this miracle can happen is obtained by noticing that the two integrands can be {\it smoothly deformed into each other} by introducing a deformation \mbox{parameter $t$}; we demonstrate $t$-independence explicitly for both NMHV and N$^2$MHV amplitudes. The equality between the objects must then be a consequence of a more general statement about amplitudes, which should follow from a simple residue theorem. We identify this simple residue theorem for all NMHV amplitudes---it is the same as the ``$\delta$-relaxing" deformation used in \cite{ArkaniHamed:2009sx} to expose the CSW recursion relations.

The outline for the paper is as follows. In the next two sections we give a general introduction to our two main themes. In \mbox{section \ref{deformation_section}} we discuss the relationship between the two different kinds of Grassmannian particle interpretations we encounter. In \mbox{section \ref{NMHV_section}} we discuss NMHV tree amplitudes. In \mbox{section \ref{NNMHV_section}} we move on to the N$^2$MHV amplitudes, and in particular, give a detailed discussion of the 8-particle N$^2$MHV amplitude. We end with brief concluding remarks in \mbox{section \ref{discussion_section}}.

\newpage
\section{Unification of Residues}\label{unification_of_residues_section}\setcounter{equation}{0}

We begin by returning to the momentum space formula for ${\cal L}_{n,k}$ given in equation (\ref{eq:mome}).
Gauge-fixing the GL$(k)$-invariance, leaves $k n-k^2=k(n-k)$ integration variables, and after imposing all $2n$ of the $\delta$-functions, we end up with an overall momentum-conserving $\delta$-function and an integral over $k(n-k)-(2n-4)=(k-2)(n-k-2)$ variables. For brevity, we will denote this total number of integration variables by $M$,
\be
M \equiv (k-2)(n-k-2),
\ee
and denote the free variables by $\tau_1, \ldots, \tau_M$. In the following, we strip-off all overall factors and concentrate on
\be
\int d^M\tau \frac{1}{(1\,2\cdots k)(2\,3\cdots k\smallplus1)\cdots (n\,1\cdots k\smallminus1)(\tau)}.
\ee
This is a holomorphic integral---i.e.\ , it is over $\tau$ and not $\bar\tau$; therefore, it must be interpreted as a contour integral in $M$ complex variables.

\subsection{Local Residues}
There is a very natural way of defining ``local residues" for functions of $M$ complex variables $\tau = (\tau_1, \ldots, \tau_M)$. Consider a rational function of the form
\be
f = \frac{g(\tau)}{p_1(\tau) p_2(\tau) \cdots p_N(\tau)}
\ee
where $N \ge M$. A residue is naturally associated with locations $\tau_*$ in $\tau$ space where $M$ of the polynomial factors $p_{i_1}(\tau_*),\ldots, p_{i_M}(\tau_*)=0$. It is natural to re-write
\be
f = \frac{h_{i_1,\ldots,i_M}(\tau)}{p_{i_1}(\tau) \cdots p_{i_M}(\tau)}\, \, \,\quad {\rm with} \quad\, \, h_{i_1, \ldots, i_M}(\tau)= \frac{g(\tau_*)}{\prod_{j \neq i_{1,\ldots,M}} p_j(\tau_*)}.
\ee
In the neighborhood of such a point we can change variables from $(p_{i_1}, \ldots, p_{i_M}) \to (u_1, \ldots, u_M)$, and up to a Jacobian, the integral becomes $\int du_1/u_1 \cdots du_M/u_M$, which is naturally defined to have residue $1$. We denote the residue as $(p_{i_1})(p_{i_2})\cdots(p_{i_M})$, given by
\be\label{eq:losa}
(p_{i_1})(p_{i_2}) \cdots (p_{i_M})|_{\tau_*} = \frac{h_{i_1, \ldots, i_M}(\tau_*)}{{\rm det}\left(\frac{\partial(p_{i_1},\ldots,p_{i_M})}{\partial(\tau_1,\ldots,\tau_M)}\right)(\tau_*)}.
\ee
%
Note that this definition of the residue depends on the order in which the polynomials enter in the Jacobian and is naturally antisymmetric in the labels: different orders can give answers which differ by a sign. This is a reflection of the fact that we were supposed to choose an orientation for the contour. The contour is in fact topologically a collection of circles \mbox{$T^m = \{\tau : |p_i(\tau)|=\epsilon_i \}$} and the orientation that produces (\ref{eq:losa}) is given by $d(\arg(p_{i_1}))\wedge\cdots \wedge d(\arg(p_{i_M}))$.

The NMHV tree amplitudes are given as a sum over these simple local residues. Consider the $n=7$ NMHV amplitude. In \cite{ArkaniHamed:2009dn}, the BCFW-contour for the amplitude was found to be given as
\be
\Gamma^{{\cal L}}_{7,3} = (2)\left[ (3)+(5)+(7)\right] + (4)\left[(5)+(7)\right] + (6)(7).
\ee
Each term is of the form $(i)(j)$ with $(i)$ representing the minor $(i\,\,i\smallplus1\,\,i\smallplus2)$. The BCFW-contour for general NMHV amplitudes is of the form
\be
\label{nmhvcont}
\Gamma^{{\cal L}}_{n,3} =\Large\sum\normalsize\underbrace{(e_1)(o_2)(e_3)\,\,\cdots\phantom{+}}_{\text{\normalsize{$n-5$ terms}}}\,\,,
\ee
where the sum is over all strictly-increasing series of $(n-5)$ alternating even ($e$) and \mbox{odd ($o$)} integers. Again, this form is not unique: as shown in \cite{ArkaniHamed:2009dn}: using residue theorems one can exchange the role of even and odd integers in this sum in many ways---and this fact was important to the proof given in \cite{ArkaniHamed:2009dn} of the cyclic-invariance of the entire contour.

For $k>3$, it is clear that for large-enough $n$, the simplistic definition of a local residue described above is inadequate to localize the integrand: we have $n$ minors, but $(k-2)(n-k-2)$ variables, which exceeds $n$ for any $k>3$ for some sufficiently-large $n$. However, as explained in more detail in \cite{ArkaniHamed:2009dn}, our object allows for a more refined notion of ``composite residue'' which is applicable when there are fewer polynomial factors than there are variables. This allows residues to be defined for any $n$ and $k$. A simple illustration of a composite residue is given by the function of three variables $x,y,z,$
\be
\frac{1}{x(x + yz)}.
\ee
Note that there are only two polynomial factors in the denominator, and so it is not possible to define a local residue in the standard way. Nonetheless, on the locus where the first polynomial factor vanishes, $x=0$, the second polynomial factorizes as $y \cdot z$, and one should reasonably define this to have residue 1. Note that such a ``composite" residue is only possible for very special functions: had we replaced the second polynomial factor with $(x + yz + a)$ for $a \neq 0$, no such identification would be possible. Geometrically, for $a=0$, the set of points where both the polynomials vanish splits into two infinite families $(x=0,y=0,z)$ and $(x=0,y,z=0)$, and the point where the residue is defined is the intersection of these infinite families. As discussed in \cite{ArkaniHamed:2009dn}, exactly the same phenomenon happens with the minors of the ${\cal L}_{n,k}$: on the zeros of some of the minors, other minors factor into pieces, each of which can be individually set to zero to define composite residues. Already for the 8-point N$^2$MHV-amplitude, some of the objects appearing the BFCW form of the tree amplitude are composite residues.
Below, we will find a very natural way of thinking about composites that is a natural consequence of our new picture for unifying residues into a single variety: composite residues can be thought of as {\it ordinary} residues, but associated with putting minors made of {\it non-consecutive} columns to zero.

\subsection{Tree Contour as a Variety}

The NMHV tree contour defined by $\Gamma^{{\cal L}}_{n,3}$ in (\ref{nmhvcont}) is perfectly clear as given. However, there is something somewhat unnatural about it: it is not precisely a ``contour" in the sense used by mathematicians. The reason is that we haven't presented
the set of residues we are summing-over as a subset of the zeros of a {\it single} mapping from $\mathbb{C}^M \to \mathbb{C}^M$; in other words, we haven't identified a fixed set of $M$ polynomials $(f_1, \ldots, f_M)$, such that the tree contour is contained in a subset of the solutions to $f_i = 0$. In fact for NMHV amplitudes it is possible to do this for $n=6,7$, taking the $f$'s to be made of products of the consecutive minors appearing in the denominator of ${\cal L}_{n,k}$. However, already for $n=8$, we'll see that it is impossible to do this using only {\it consecutive} minors. Thus, we seem to reach an impasse: from a mathematical point of view, it would clearly be natural to ``glue" all the residues together as zeros of a single map---to think of the contour as a single algebraic variety. But the physical contour for tree amplitudes does not seem to admit such an interpretation.

However, we will see that it {\it is} possible to naturally unify the residues into a single variety---the apparent obstruction to doing so was merely a consequence of the myopia of only considering minors composed of consecutive columns of $C_{\alpha a}$. 

By iteratively adding one particle at a time, we will soon see that the tree-level amplitude {\it can} be given in the form
\be
\int\limits_{\vf=0} d^M\tau \frac{h(\tau)}{f_1(\tau)\ldots f_M(\tau)},
\ee
where we sum over all the zeros of $\vf\equiv(f_1,\ldots,f_M)=0$. Note that $h(\tau)$ is not just a polynomial, but a ratio of polynomials---otherwise this sum would vanish by the global residue theorem! The remarkable fact is that, as rational functions,
\be
\label{newform}
\frac{h}{f_1\cdots f_m} = \frac{1}{(1\,2\cdots k)(2\,3\cdots k\smallplus1)\cdots (n\,1\cdots k\smallminus1)},
\ee
but the numerator of $h$ and $f_1,\ldots ,f_M$ are polynomials {\it in the minors} of $C_{\alpha\,a}$ of degree larger \mbox{than $n$}, and all the non-consecutive minors appearing in the $f_i$'s are cancelled by those in the numerator of $h$. This is how they manage to encode the information about the contour.

For instance, we will show that {\it all} NMHV amplitudes can be written in the form
\be
\label{firstnmhv}
A^{(3)}_n = \int\limits_{\vf_n=0} \frac{\prod_{j=6}^{n-1}[(1\,\,2\,\, j)(2\,\, 3\,\,j\smallminus1)]}{(n\smallminus1)(1)(3)\, f_6\cdot f_7\cdots f_{n}},
\ee
where $\vf_n=(f_6,\ldots,f_n)$ and 
each $f_k:\mathbb{C}\to \mathbb{C}$ is given by the product of minors,
\be
f_k=(k\smallminus2\,\,k\smallminus1\,\,k)(k\,\,1\,\,2)(2\,\,3\,\,k\smallminus2).
\ee
Similarly, each N$^2$MHV amplitude can be written as
\be
\label{firstnnmhv}
A_n^{(4)}=\int\limits_{\vf_n=0} \frac{\prod_{j=7}^{n-1}\big[\left(1\,2\,3\,j\right)\left(2\,3\,j\smallminus2\,j\smallminus1\right)\left(1\,j\smallminus2\,j\smallminus1\,j\right)\big]\prod_{j=4}^{n-3}\big[\left(1\,3\,j\,j\smallplus1\right)\left(1\,2\,j\,j\smallplus3\right)\big]}{(n\smallminus1)(1)(3)\quad\mathscr{F}_7\!\cdot \! \mathscr{F}_8\cdots \mathscr{F}_n},
\ee
where $\vf_n\equiv\left(f_{7_a},f_{7_b},f_{8_a},f_{8_b},\ldots,f_{n_a},f_{n_b}\right)$ with \be\begin{split}
&f_{\ell_a}\equiv(\ell\smallminus3\,\,\ell\smallminus2\,\,\ell\smallminus1\,\,\ell)(\ell\smallminus3\,\,\ell\,\,1\,\,2)(\ell\smallminus3\,\,2\,\,3\,\,\ell\smallminus2);\\
\mathrm{and\,\,}\qquad&f_{\ell_b}\equiv(1\,\,\ell\smallminus2\,\,\ell\smallminus1\,\,\ell)(1\,\,\ell\,\,2\,\,3)(1\,\,3\,\,\ell\smallminus3\,\,\ell\smallminus2);
\end{split}
\ee
and for which $\mathscr{F}_{\ell}\equiv f_{\ell_a}\cdot f_{\ell_b}$.

Note that as stated the definitions of $h$ and $f$ include minors built out of {\it non-consecutive} columns. We will see that their presence is crucial for allowing us to unify all the residues into a single algebraic-variety. As a by-product, they will also teach us how to think about ``ordinary" and ``composite" residues of ${\cal L}_{n,k}$ in a more uniform way, as ``composite" residues can be understood as ordinary residues involving non-consecutive minors.

\newpage

\subsection{Manifest Soft-Limits and the Particle Interpretation}
We motivated the gluing-together of tree-amplitude residues into a
single variety from a mathematical point of view. There is also a
physical reason to be dissatisfied with the usual way of presenting
tree-amplitudes as a sum over disparate local residues: soft-limits of the amplitude would then not then manifest themselves as an obvious feature of the contour. Suppose we take the holomorphic
soft-limit of particle $n$, where $\lambda_n\to 0$ while keeping $\tilde\lambda_n$ fixed. In this limit, the most singular part
of the amplitude connects directly to the lower point amplitude with the usual multiplicative soft factor
\be \label{eq:softy} A_{n} \to \frac{\langle
n\smallminus1~1\rangle}{\langle n\smallminus1~ n \rangle \langle n~ 1\rangle}A_{n-1}. \ee This means that there must be a connection between
$\Gamma^{{\cal L}}_{n,k}$ and $\Gamma_{n-1,k}^{{\cal L}}$; but this
is not at all manifest for the NMHV tree contour given by
equation (\ref{nmhvcont}). It is important to mention that from the
mathematical point of view, the {\it inverse}  operation is in fact more
natural. In other words, it is more natural to think about the
inclusion of $G(k,n-1)$ into $G(k,n)$ than to think about the
projection of some contour in $G(k,n)$ down to $G(k,n-1)$. Indeed,
in \cite{ABCCK:2010}, we will show that there is a natural notion of
an ``inverse-soft" operation on individual residues, that maps a residue of ${\cal L}_{n,k-1}$ to a residue of ${\cal L}_{n,k}$.  However what we
are after here is a remarkable feature not of individual residues
but of the way they are combined into $\Gamma^{{\cal L}}_{n,k}$.

Quite beautifully, the unification of residues in equation (\ref{newform}) allows us to think of the $n$-particle amplitude by ``adding a particle" to the $(n-1)$-particle amplitude in a way that makes the soft-limits manifest. In fact, we can write
\be
\frac{h_n}{f_1 \cdots f_{M_n}} = \frac{h_{n-1}}{f_1 \cdots f_{M_{n-1}}} \times \inversesoft{(n-1) \to n}{}
\ee
and recursively build the contour for higher point amplitudes in this way. Furthermore, in the soft limit, $\lambda_n \to 0$, we find that (after an application of the global residue theorem) the $\tau$ integral localizes so that
\be
\inversesoft{(n-1)\to n}{} \to \frac{\langle n\smallminus1~1\rangle}{\langle n\smallminus1~ n \rangle \langle n~ 1\rangle},
\ee
which precisely reproduces the needed soft factor!

\newpage
\subsection{Connection to CSW Localization}

The attentive reader may have noticed that the forms of $f_i$ presented above for the NMHV and N$^2$MHV amplitudes contain the product of three minors; moreover the denominator of $h_n$ is the product of the three consecutive minors $(n\smallminus1),(1)$ and $(3)$. This is not an accident: these forms are intimately connected to localization of amplitudes on CSW configurations in twistor space! In order to understand why, let us begin by noting that it is natural to think of the matrix $C_{\alpha a}$ as a collection of $n$ $k$-vectors, or $n$ points in $\mathbb{C}^k$. In fact, due to the little group symmetry which rescales each column of $C_{\alpha\,a}$ independently, we can think of these points projectively as $n$ points in $\mathbb{CP}^{k-1}$. Since the contour of integration is the variety where $\vf= 0$, it is natural to ask whether there is anything special about the points in  $\mathbb{CP}^{k-1}$ for which $\vf$ vanishes? In fact, there is an even more interesting question, which we can best discuss with some new notation. Let us define the ``expectation value" of some ``operator" built out of minors of $C_{\alpha\,a}$, by
\be
\langle \mathcal{O} \rangle = \int\limits_{\vf=0} \frac{h}{f_1 \cdots f_M} {\cal O}\,\,.
\ee
Note that with this definition, the amplitude itself is $\langle 1 \rangle$, and trivially $\langle f_i \rangle = 0$. However there are also other operators with vanishing expectation values. For instance, taking the operator to be the denominator of $h_n$, we find that $\langle (n\smallminus1) (1) (3) \rangle = 0$ as a consequence of the global residue theorem. One might ask whether there exists a different way of writing the integral where all these vanishing expectation values are understood on the same footing trivially, as part of the definition of the contour of integration. In this case the answer is ``yes": the ``$\delta$-relaxing" contour-deformation used in \cite{ArkaniHamed:2009sx} does this. We see that this form of the amplitude makes a certain localization property of the amplitude manifest---associated with the vanishing ``expectation value" of objects built out of the product of three minors. If we further use the (independently proven) information that the amplitude is cyclically invariant, we get a very large number of constraints, which we can loosely think of as localizing the integral in the Grassmannian.

Now, for $k \leq 4$, there is a very close connection between {\it localization in the Grassmannian} and {\it localization in (Z) twistor space}. In order to see this, it suffices to Fourier-transform the bosonic parts of the kinematical $\delta$-functions $\delta^{4|4}(C_{\alpha a}{\cal W}_a)$ into the $Z$ twistor space:
\be
\label{wtoz}
\prod_\alpha \delta^{4}(C_{\alpha a} {\cal W}_a) \to \int d^{4} z^{\alpha} \prod_a \delta^{4}(Z_a - C_{\alpha a} z^\alpha).
\ee
Note that for $k=3$, the twistor space ``collinearity operator" $\epsilon_{IJKL} Z^I_i Z^J_j Z^K_k$ acts on the amplitude \nolinebreak as 
\be 
\label{twistloc}
(Z_i Z_j Z_k)^I A_{n} = \int d^{4} z (z \, z \, z)^I \langle (i \, j \, k) \rangle\,\,.
\ee 
We can think of the  ``localization in the Grassmannian" implied by $\langle (i \, j \, k \,) \rangle =0$ as telling us that the points $\{i, j, k\}$ in the $\mathbb{CP}^2$ associated with the columns of $G(3,n)$ are (projectively) collinear. By virtue of equation~(\ref{twistloc}) this tells us that this sense of localization in the Grassmannian is sharply reflected as localization in twistor space.

All of this is interesting because the set of twistor space collinearity operators that test for CSW localization precisely involve products of {\it three} of them---which translate to the vanishing expectation value for the product of three minors in the Grassmannian. 
It is very easy to see that for any configuration of $n$ cyclically ordered points localized on two lines in $\mathbb{CP}^2$, the product of three minors $(i \,  x \,  j)(k \, y \, l)(m \, z \, o)$ vanishes, where $i<x<j \leq k < y < l \leq m < z < o$. To prove it, let's assume that the first two factors are not equal to zero, which means that $(i\,x\,j),(k\,y\,l)$ can not be collinear. This forces the points to be distributed on the two lines as in:
\begin{figure}[h]\vspace{-0.4cm}
\centering\includegraphics[scale=.6]{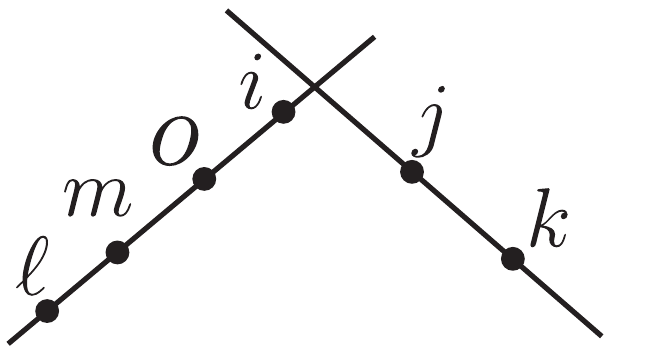}\vspace{-0.4cm}
\end{figure}

\noindent But then $m,z,o$ are forced to be on the same line, and so the last factor $(m\,z\,o)=0$. This shows why two minors are insufficient but three suffice. Furthermore, having sufficiently many of the operators of this form vanish is enough to {\it guarantee} CSW-localization. 
Something similar is true for $k=4$. Here the coplanarity operator $(Z_i Z_j Z_k Z_l)$ in twistor space maps to the $4 \times 4$ minor $(i \, j \, k \, l)$ in the Grassmannian. Perhaps a little surprisingly, collections of coplanarity operators suffice to ensure CSW-localization on lines. This can happen if the coplanarity conditions involve non-consecutive points. 

For $k>4$, it is in general difficult to find a set operators testing localization for CSW configurations of $(k-1)$ intersecting lines in the $\mathbb{CP}^3$ of twistor space; the reason is that the $\mathbb{CP}^3$ is too ``small". It is however much easier to talk about localization to CSW-like configurations of $(k-1)$ lines in $\mathbb{CP}^{k-1}$, and this is precisely the natural question associated with vanishing operator expectation values from the Grassmannian point of view! It is amusing to ask what ``Grassmaniann CSW" operators test for this Grassmannian notion of localization. It is easy to exhibit two large classes of such operators, always made from the products of three minors for any $k$. One class is similar to set we described for $k=3$: the product of three $(k \times k)$ minors $(i \cdots j)(k\cdots l)(m \cdots n)$ vanishes for  CSW-like configurations in $\mathbb{CP}^{k-1}$. Another class of operators can be easily constructed recursively. Given any configuration localized on lines in $\mathbb{CP}^{k-1}$, we can project down along one of the lines to get a another set of points (with some co-incident) localized on $(k-2)$ lines in $\mathbb{CP}^{k-2}$, as shown below in an example with $k=4$:
\begin{figure}[h]\vspace{-0.4cm}
\centering\includegraphics[scale=.6]{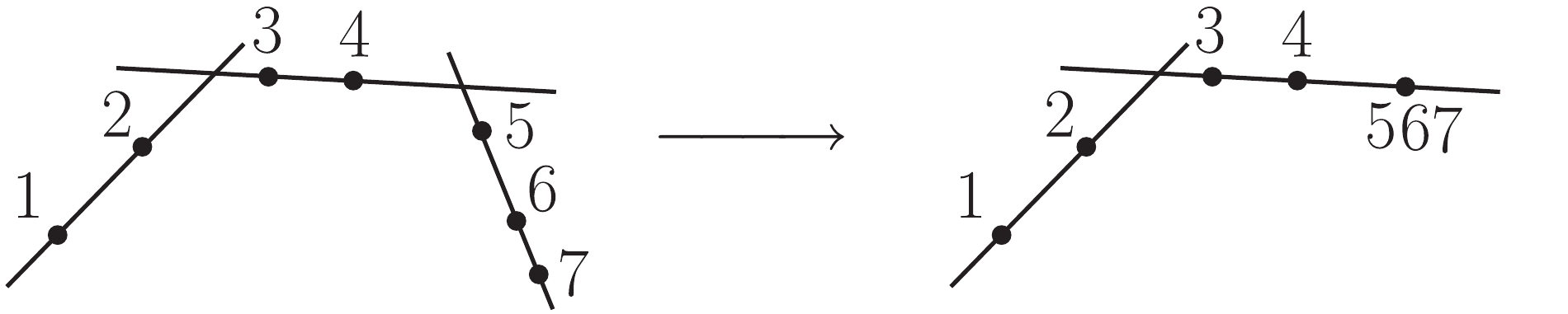}\vspace{-0.4cm}
\end{figure}

\noindent Since any particle $I$ belongs to a unique line, by considering $(k \times k)$ minors that all include $I$, we are projecting-down along the line containing $I$ to the problem in $\mathbb{CP}^{k-2}$. Thus the set of operators obtained by attaching column $I$ to the ones just discussed---of the form $(I \,i \cdots j)(I \,k \cdots l)(I\,m \cdots o)$---will also vanish on these configurations. Given that localization to ``Grassmannian" CSW configurations implies localization on CSW configurations in twistor space, this strongly suggests that this ``three-minor" form of the maps obtained in unifying tree amplitudes should persist for all $k$. 

A very non-trivial check on this picture can be made by examining the simplest amplitude with $k=5$---the split helicity 10-particle amplitude.
There are 20 different BCFW terms in the amplitude,  which can all be easily identified as residues of ${\cal L}_{10,5}$. We can test for localization in the Grassmannian by computing $\langle \mathcal{O}_{{\rm CSW}} \rangle$ for the class of Grassmannian CSW operators we have just defined. Since we know the form of the $C$-matrix explicitly for each residue, this simply amounts to taking each BCFW term and multiplying it by the relevant product of three minors of its associated $C$-matrix. We have checked that the correct linear combination of twenty BCFW terms weighted with ${\cal O}_{{\rm CSW}}$ in this way indeed vanishes. Something even stronger is true: we checked that if we leave the coefficients of all 20 BCFW terms arbitrary,  demanding that all the ``localization on intersecting lines in $\mathbb{CP}^4$" operators annihilate the amplitude completely fixes the 20 terms up to a single overall scale. We will return to further investigate these fascinating issues at greater length in a future work.

\section{Veronese Particle Interpretation}\label{particle_interpretation_section}\setcounter{equation}{0}


In the previous section, we discovered the particle interpretation and CSW localization of the tree amplitudes as a
happy consequence of gluing together the
residues of ${\cal L}_{n,k}$ contributing to the tree amplitude into a single variety. But the particle interpretation was not manifest from the outset---nor was the cyclic-invariance of the amplitude.

This motivates us to start anew, and construct a
Grassmannian theory which makes the particle interpretation and cyclic-symmetry as manifest as possible. We will find that this
straightforward exercise leads us essentially uniquely to the
connected prescription \cite{Roiban:2004yf} of  Witten's twistor
string theory \cite{Witten:2003nn}. As an additional bonus, in
addition to cyclic symmetry, this formulation will make the famous
$U(1)$-decoupling identity manifest, which is a remarkable property of
amplitudes that is only ``obvious" from the Lagrangian point of
view.

Going back to the beginning, the central object encoding ``Grassmannian kinematics" are the twistor-space $\delta$-functions which contain the only dependence on space-time variables $\prod_\alpha \delta^{4|4}(C_{\alpha a} {\cal W}_a)$. As seen recently in \cite{Bullimore:2009cb, Kaplan:2009mh}, this factor alone goes a long way in explaining how the (non-trivial) kinematics of leading singularities can be encoded in ${\cal L}_{n,k}$, even without using any specific properties of the measure made from consecutive minors, so clearly we should stick with this structure. Transforming back to momentum space it becomes
\be
\prod_\alpha \delta^2(C_{\alpha a} \tilde \lambda_a) \delta^4(C_{\alpha a} \tilde \eta_a) \int d^{2 \times k} \rho^\alpha \prod_a \delta^2(\rho^\alpha C_{\alpha a} - \lambda_a).
\ee
The bosonic $\delta$-functions impose $(2n - 4)$ constraints on $C_{\alpha a}$, enforcing the geometric constraint that the $k$-plane $C_{\alpha\,a}$ by orthogonal to the 2-plane $\tilde \lambda$ and contains the 2-plane $\lambda$.
Now, in equation (\ref{eq:mome}), in interpreting the integral over $G(k,n)$ as a contour integral, we place a further $(k-2) \times (n- k -2)$ constraints on $C_{\alpha a}$, which is equivalent to declaring that we are performing the integral over a $k \times (n-k) - (k-2)\times(n-k-2) = (2n - 4)$-dimensional sub-manifold in $G(k,n)$. We can generalize this idea to define a whole class of ``Grassmannian theories", which enforce the  ``kinematic" constraints on the space-time variables associated with $\delta^{4|4}(C_{\alpha a} {\cal W}_a)$. We simply choose some $(2n - 4)$ dimensional subspace $\Sigma$ of the Grassmannian, a general point of which we represent as $C^{\star}_{\alpha a}(\zeta_I)$ for $I = 1, \ldots, (2n - 4)$. Then we consider the object
\be
\int_\Sigma d^{2n - 4} \zeta \, \mu(\zeta) \prod_\alpha \delta^{4|4}(C^{\star}_{\alpha a}(\zeta_I){\cal W}_a),
\ee
where $\mu(\zeta)$ is a measure factor.

Now, of all such Grassmannian theories, there is a special class that we can motivate physically as having a ``particle interpretation". Ordinarily, the configuration space for $n$-particles is thought of as $n$ copies of a given space on which each of the particles ``live". In order for a Grassmannian theory to have such a ``particle interpretation", then, we would like to loosely think of $\Sigma = (\Sigma_{{\rm base}})^n$. Now, $\dim\left(\Sigma\right) = (2n - 4)$ (let us leave the $-4$ offset for a moment, and) note that at large $n$, the only way we can make such an identification is if $\dim\left(\Sigma_{{\rm base}}\right)$ = 2; and so the most natural choice is $\Sigma_{{\rm base}} = \mathbb{C}^2$. The ``$-4$" can arise from a GL(2)-redundancy acting on $\mathbb{C}^2$. We can therefore conclude that we are looking for a $(2n-4)$ sub-manifold of the Grassmannian, that can be thought of as a mapping of $(\mathbb{C}^2)^n$/GL(2) into $G(k,n)$. It only remains to discuss how to determine this mapping from $(\mathbb{C}^2)^n$/GL(2)$\to G(k,n)$ explicitly.

Let us denote a general point in $\mathbb{C}^2$ by $\sigma = (A, B)$. It is natural to look for a mapping into a point we will denote by $\sigma^V(\sigma)$ in $\mathbb{C}^k$, such that the GL(2)-action on $\sigma$ turns into some GL($k$)-action on $\sigma^V$. There is a canonical map from $\mathbb{C}^2 \to \mathbb{C}^k$, familiar from elementary algebraic geometry which does this precisely and is known as the Veronese map:
\be
\sigma :\left(\begin{array}{c} A \\ B \end{array} \right) \to \left(\begin{array}{c} A^{k-1} \\ A^{k-2} B \\ \vdots \\ B^{k-1} \end{array} \right) \equiv \sigma^V(\sigma).
\ee
We can assemble the $n$ $k$-dimensional vectors $\sigma_a^V$, for $a=1, \ldots, n$, into the $k \times n$ dimensional matrix $C^V_{\alpha a}[\sigma]$ which denotes the Veronese map from $(\mathbb{C}^2)^n/$GL(2)$\to G(k,n)$
\be
\label{veronese}
C^V[\sigma] = \left(\begin{array}{cccc} \vdots & \vdots & \cdots & \vdots \\ \sigma^V[\sigma_1] & \sigma^V[\sigma_2] & \cdots & \sigma^V[\sigma_n] \\ \vdots &  \vdots & \cdots &\vdots \end{array} \right);
\ee
 or written more succinctly
 \be
 C^V_{\alpha a}[\sigma] = A^{k-\alpha}_a B^{\alpha-1}_a.
 \ee
We group all the $\sigma_a$ together into $2 \times n$ matrix which, given the GL(2)-action, we can think of as an element of $G(2,n)$. Thus we can also think of $C^V$ as giving the Veronese map from $G(2,n) \to G(k,n)$.

\subsection{Twistor String Theory}

In order to complete our story and fully define a Grassmannian theory, we need to integrate over the two-dimensional vectors $\sigma_a$ with a natural GL(2)-invariant measure. By analogy with the simple choice for the GL($k$)-invariant measure chosen in equation (\ref{eq:mome}), the simplest possibility is to soak-up the GL(2) weights with a product of consecutive $2 \times 2$ minors and define
\be
 \label{connected}
 {\cal T}_{n,k}({\cal W}) = \frac{1}{{\rm vol(GL}(2))} \int \frac{d^2 \sigma_1 \cdots d^2 \sigma_n}{(\sigma_1 \sigma_2)(\sigma_2 \sigma_3) \cdots (\sigma_n \sigma_1)} \prod_\alpha \delta^{4|4}(C^V_{\alpha a}[\sigma] {\cal W}_a).
\ee
In the case of equation (\ref{eq:mome}) for ${\cal L}_{n,k}$,  the choice of measure with consecutive minors had much more than aesthetic benefits: only with this choice was it possible to prove the equivalence with equation (\ref{second}) and establish dual superconformal invariance. Similarly, in the present case, the choice of measure with the product of the $(\sigma_i \sigma_{i+1})$ in the denominator makes a remarkable feature of scattering amplitudes manifest which is normally only obvious from the spacetime Lagrangian. This property is the famous ``$U(1)$-decoupling identity".
While we normally talk about color-stripped amplitudes, in reality the full amplitude is given by a sum over permutations
\be
{\cal A}_n = \sum_{P \in S_n/{\mathbb{Z}_n}}{\rm Tr}\left( T^{a_{P(1)}}T^{a_{P(2)}}\cdots T^{a_{P(n)}}\right) A(P(1),\ldots, P(n)).
\ee
When the gauge group is taken to be any product of $SU(N_i)$ factors (including $U(1)$'s), the Lagrangian description makes it obvious that the amplitude for producing particles in the adjoint of $SU(N_i)$ from $SU(N_j)$-particles must vanish. This implies many relations among the partial amplitudes $A(P(1),\ldots, P(n))$ with different orderings. The simplest of these relations is called the $U(1)$-decoupling identity, which is obtained when the gauge group is taken to be $U(N)=U(1)\times SU(N)$. Now, the dependence on the external spacetime variables in $\delta^{4|4}(C^V_{\alpha a}[\sigma] {\cal W}_a)$ is fully permutation-invariant; the only factor that breaks the permutation invariance
down to cyclic invariance is the factor $(\sigma_1 \sigma_2)(\sigma_2 \sigma_3) \cdots (\sigma_n \sigma_1)$, and it is trivial to see that this satisfies the identity necessary for ${\cal T}_{n,k}({\cal W}_a)$ to satisfy the $U(1)$-decoupling identity.

We have motivated equation (\ref{connected}) as a beautiful way of writing a theory enforcing a Grassmannian ``particle interpretation". It is also nothing other than the connected prescription \cite{Roiban:2004yf} for Witten's twistor string theory \cite{Witten:2003nn} (see also \cite{Mason:2009sa} where the Grassmannian form of the twistor string theory is presented). To see this, we Fourier-transform from the ${\cal W}_a$ to the ${\cal Z}_a$ variables in order to return to Witten's original setting:
\be
\prod_\alpha \delta^{4|4}(C^V[\sigma]_{\alpha a} {\cal W}_a) \rightarrow \int d^{4|4} z^{(\alpha)} \prod_a \delta^{4|4}({\cal Z}_a - C^V[\sigma]_{\alpha a} z^{(\alpha)}).
\ee
If we further write $\sigma_a = (A_a \,  B_a) = \xi_a (1 \, \rho_a)$, the GL(2) action has a GL(1) rescaling the $\xi$ and an SL(2) acting on $\rho$, with $(1 \,  \rho)$ being thought of as inhomogeneous co-ordinates on $\mathbb{CP}^1$. Then, $(\sigma_i \sigma_{i+1}) = (\xi_i \xi_{i+1}) (\rho_i - \rho_{i+1})$, and we have
\be
\label{connected2}
{\cal T}_{n,k}({\cal Z}_a) = \frac{1}{{\rm vol(GL(2))}} \int \frac{dz^{(\alpha)}\;\;\;d \rho_1 \cdots d \rho_n}{(\rho_1 - \rho_2)(\rho_2 - \rho_3) \cdots (\rho_n - \rho_1)} \prod_a \delta^{3|4}({\cal Z}_a - \sum_{\alpha=0}^{k-1} z^{(\alpha)} \rho_a^{\alpha}),
\ee
where $\delta^{3|4}({\cal Z} - {\cal Z}^\prime)$ is a projective $\delta$-function in $\mathbb{CP}^{3|4}$:
\be
\delta^{3|4}({\cal Z} - {\cal Z}^\prime) = \int \frac{d \xi}{\xi} \delta^{4|4}\left( {\cal Z} - \xi {\cal Z}^\prime \right).
\ee
Equation (\ref{connected2}) is exactly the connected prescription for computing tree amplitudes from twistor string theory, integrating over the moduli space (parametrized by the $z^{(\alpha)})$ of degree-$(k-1)$ curves in $\mathbb{CP}^{3|4}$. However, notice that from the point of view of the Grassmannian, there is a more fundamental notion of localization: under the action of the little group, ${\cal W}_a \to t_a {\cal W}_a$, we have $C_{\alpha a} \to t_a^{-1} C_{\alpha a}$, and therefore we can think of each column of $C_{\alpha a}$ projectively as giving a point in $\mathbb{CP}^{k-1}$. The Veronese condition of equation (\ref{veronese}) is then nothing but the statement that all these points in $\mathbb{CP}^{k-1}$ lie on a degree-$(k-1)$ mapping of $\mathbb{CP}^1 \to \mathbb{CP}^{k-1}$. This localization to degree-$(k-1)$ curves in $\mathbb{CP}^{k-1}$ associated with the Grassmannian implies, via equation (\ref{connected2}), localization on degree-$(k-1)$ curves in twistor space.

We can cast the expression for ${\cal T}_{n,k}$ in a form that will
most directly facilitate a comparison with ${\cal L}_{n,k}$, by
writing ${\cal T}_{n,k}$ as an integral over the full Grassmannian
$G(k,n)$, with \mbox{$(k-2) \times (n-k-2)$} $\delta$-functions imposing
the constraint that the $k$-planes have the Veronese form of
equation (\ref{veronese}) with a ``particle interpretation". We do
this by formally introducing ``1" in the form \be 1 = \frac{1}{{\rm
vol(GL(}k))}\int d^{k \times n} C_{\alpha a} d^{k \times
k}L_{\alpha}^{\beta} ({\rm det}L)^n \prod_{\alpha,
a}\delta(C_{\alpha a} - L_{\alpha}^\beta C^V_{\beta a}[\sigma]); \ee
here the integral over $L_\alpha^\beta$ is just one over all $k
\times k$ linear transformations, and by gauge-fixing to
$L_\alpha^\beta = \delta_\alpha^\beta$, we get ``1" trivially.

We can then integrate over the $\sigma_a$, and we are left with \be {\cal
T}_{n,k}({\cal W}_a) = \frac{1}{{\rm vol(GL(}k))} \int d^{k \times n}
C_{\alpha a} F(C) \delta^{4|4}(C_{\alpha a} {\cal W}_a), \ee where
\be \label{fofc} F(C) = \frac{1}{{\rm vol(GL(2))}} \int \frac{d^2 \sigma_1 \cdots
d^2 \sigma_n}{(\sigma_1 \sigma_2)(\sigma_2 \sigma_3) \cdots
(\sigma_n \sigma_1)} d^{k \times k} L_{\alpha}^\beta \prod_{\alpha,
a} \delta(C_{\alpha a} - L_\alpha^\beta C^V_{\beta a}[\sigma]). \ee
Clearly, by construction $F(C)$ will contain $(k-2) \times (n-k-2)$
$\delta$-function factors localizing the integral over the $C$'s to
have the Veronese form. Really these $\delta$-functions are to be
thought of holomorphically, in other words, we think of ``$\delta(x)\to
1/x$'', where the contour of integration is forced to  enclose $x=0$
(see \cite{ArkaniHamed:2009sx}). Therefore, ${\cal T}_{n,k}$
will have the form \be \label{Seqn} {\cal T}_{n,k} = \frac{1}{{\rm
vol(GL(}k))} \int\limits_{S_1 = \cdots = S_{M} = 0} d^{k \times
n} C_{\alpha a} \frac{H(C)}{S_1(C) \cdots S_{M}(C)}. \ee
We will call the $S(C)$'s ``Veronese operators", whose vanishing
is necessary for the matrix $C_{\alpha\,a}$ to be put into the Veronese form by
some GL$(k)$ transformation.

The first non-trivial example to study is the six-particle NMHV amplitude
$n=6,k=3$; the computation was
first presented in \cite{Dolan:2009wf,Spradlin:2009qr}, having
gauge-fixed the GL($k$)-symmetry on the $C$'s in the ``link
representation" where $k$ of the columns of $C_{\alpha\,a}$ are set to an
orthonormal basis; it is very easy to translate these results in a
general GL($k$) invariant form, as has also been recently done in
\cite{Nandan:2009cc}. The result for $H(C)$ is \be H(C)
=\frac{(1\,3\,5)}{(1\,2\,3)(3\,4\,5)(5\,6\,1)} \ee while there is a single $S(C)$
given by \be S(C) \equiv S_{123456}(C) = (1\,2\,3)(3\,4\,5)(5\,6\,1)(2\,4\,6) -
(2\,3\,4)(4\,5\,6)(6\,1\,2)(3\,5\,1). \ee

\subsection{Veronese Operators for Conics}

The object $S_{123456}(C)$ will play a fundamental role in the story of the connected prescription, so we pause to discuss its salient properties. For $n=6,k=3$, the Veronese condition is simply that 6 points on $\mathbb{CP}^2$ lie on a conic. Now, any 5 generic points determine a conic, and there is clearly a single constraint for a 6$^{\rm th}$ additional point to lie on the conic determined by the first 5; this is what $S_{123455} = 0$ imposes. We can see that this is the constraint by looking at the form of the $C^V$ matrix
\be
C^V = \left(\begin{array}{ccc} 1 & \cdots & 1 \\ \rho_1 & \cdots & \rho_6 \\ \rho_1^2 & \cdots & \rho_6^2 \end{array} \right),
\ee
where we have used the little group freedom to rescale the elements of the first row to all \mbox{be 1}. Clearly, the Veronese condition should be GL($k$)-invariant, and hence we are looking for a relationship between the minors of $C_{\alpha\,a}$ that is a consequence of this special form. Note any $3 \times 3$ matrix made from columns of $C_{\alpha\,a}$ has the Vandermonde form and so the minors $(i\,\,j\,\,k)$ are very simple: $(i\,\,j\,\,k) = (\rho_i - \rho_j)(\rho_j - \rho_k)(\rho_k - \rho_i)$. In order to discover the relationship between minors implied by the Veronese condition in this case, examine the ``star of David" figure below:
\begin{figure}[h]\vspace{-0.4cm}
\centering\includegraphics[scale=1.5]{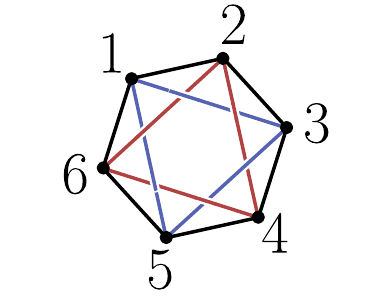}\vspace{-0.4cm}
\end{figure}

\noindent Each link in the figure connecting $(i\,j)$ represents a factor of $(\rho_i - \rho_j)$ (in cyclic order). We can interpret the product of the links $(1\,2)(2\,3)(1\,3)$ in the figure as the minor $-(1\,2\,3)$, the product $(3\,4)(4\,5)(3\,5)$ as $-(3\,4\,5)$, the product $(5\,6)(6\,1)(5\,1)$ as $-(5\,6\,1)$, and the remaining links $(2\,4)(4\,6)(2\,6) = - (2\,4\,6)$. Thus the product of all the links in the figure is $(1\,2\,3)(3\,4\,5)(5\,6\,1)(2\,4\,6)$. However the picture is clearly cyclically invariant, so the product is also $(2\,3\,4)(4\,5\,6)(6\,1\,2)(1\,3\,5)$, and thus we have found the single relation we are looking for
\be
S_{123456} = (1\,2\,3)(3\,4\,5)(5\,6\,1)(2\,4\,6) - (2\,3\,4)(4\,5\,6)(6\,1\,2)(3\,5\,1) = 0.
\ee
Clearly the condition that 6 points lie on a conic is invariant under the permutation of the points, so that if $S_{123456}=0$, then $S_{P(1) P(2) \cdots P(6)}=0$ as well.  In fact something even stronger is true. Even though it is not manifest, the object $S_{123456}$ is permutation invariant in its labels (up to the sign of the order of the permutation); in other words,
\be
S_{P(1) P(2) \cdots P(6)} = (-1)^{P} S_{12 \cdots 6}.
\ee
It is trivial to see that $S$ picks up a minus sign under a cyclic shift of the labels $i \to i+1$, and it can be further checked that $S_{123456} = - S_{213456}$ as a simple consequence of the Schouten identity.

Let us move on to examine the 7-particle NMHV amplitude
\cite{Dolan:2009wf,Nandan:2009cc,Spradlin:2009qr} where the
integrand for ${\cal T}$ is of the form \be \frac{H(C)}{S_{123456}\,\,
S_{123567}}
\ee
with
\be
 H(C) =
\frac{(1\,3\,5)(6\,1\,2)(1\,3\,6)(2\,3\,5)}{(6\,7\,1)(1\,2\,3)(3\,4\,5)}. \ee
Here the role
of the two $S$'s in the denominator is clear. The 5 points $\left\{1,2,3,5,6\right\}$
determine a conic; $S_{123456}=0$ enforces that the point $4$ lies
on this conic, while $S_{123567}=0$ enforces that $7$ lies on this
conic; together they impose that all 7 points lie on the same conic.
Actually there is a loophole in this argument, which nicely explains
the role of the many factors in the numerator of $H(C)$. If
the points $\{1,2,3,5,6\}$ lie on a {\it degenerate} conic, it is possible for both $S$'s to
vanish without having all 7 points on conic. For instance, suppose
that any four of the points $\{1,2,3,5,6\}$ are collinear; this would make each
$S$ vanish trivially, even if the other three points are in general
positions, for instance,
\begin{figure}[h]
\centering\includegraphics[scale=1]{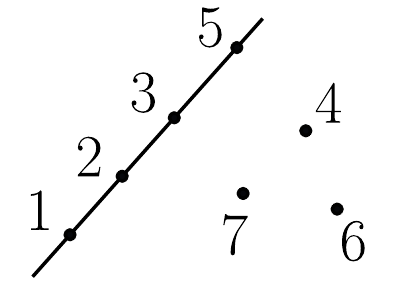}\vspace{-0.4cm}
\end{figure}

\noindent The numerator factors in $H(C)$ vanish on these ``spurious" configurations and ensure that they don't contribute to the integrand; in this example, this configuration is killed by the $(2\,3\,5)$ factor in the numerator of $H$. It is easy to check that {\it all} spurious solutions are dispatched by factors in the numerator in this way.

For general NMHV amplitudes, we will have $(n-5)$ $S$'s. We stress that there are many equivalent ways of writing equation (\ref{Seqn}), using different collections of $(n-5)$ Veronese operators in the denominator to enforce that the $n$ points lie on a conic. For instance, one canonical choice involves using a fixed set of $5$ points $\{1,2,3,4,5\}$ to determine the conic, and then simply choosing the $(n-5)$ $S$'s to be $S_{12345j}$ for $j=6, \ldots, n$. However, this is not the only possibility; all that is needed is for the labels of the $S$'s to overlap sufficiently to guarantee all $n$ points to lie on the same conic; but we will find other choices to be more natural for our purposes.

\subsection{General Veronese Operators}

Moving beyond NMHV amplitudes,  we must encounter Veronese operators
that enforce $n$ points to live on a degree-$(k-1)$ curve in
$\mathbb{CP}^{k-1}$. The conditions must again be GL$(k)$-invariant
and must therefore be written in terms of $k \times k$ minors.
Fortunately, it is very easy to see that the conditions are always
a collection of constraints of exactly the same form as $S_{123456} = 0$, involving the
difference of the product of 4 minors. Physically this is because we can use parity to relate the Veronese conditions for $(n,k)$ to those
for $(n,n-k)$. It is illuminating to see this explicitly, since it
also allows us to make contact with the work of \cite{Dolan:2009wf}.
Parity is manifest in the link representation, so let us study what
the Veronese $C^V$ matrices look like in this representation. Suppose
we gauge-fix the first $k$ columns to the $k \times k$ identity
matrix, and denote the remaining entries as $c_{iI}$ for $i=1,
\ldots, k$ and $I= k+1, \ldots, n$. Instead of finding the explicit
GL($k$) transformation that takes the $C^V$ matrix to this form, we
can note that the $c_{iJ}$ can be written in a GL$(k)$ invariant way
as the ratio of two minors: \be \label{PL} c_{iI} = \frac{(1 2
\cdots \hat{i} \cdots k I)}{(12 \cdots k)}, \ee where in the
numerator $\hat{i}$ denotes that the column $i$ is not included.
Since this ratio is GL($k$)-invariant, we can compute it directly
for the form $C^V$, easily finding \be c_{iI} =
\frac{\kappa_I}{\kappa_i} \frac{1}{\rho_I - \rho_i} \ee where \be
\kappa_I = \prod_{j=1}^k (\rho_I - \rho_j), \, \, \kappa_i =
\prod_{j \neq i=1}^k (\rho_i - \rho_j). \ee So the Veronese operators
must check whether the $k\times(n-k)$ variables $c_{iI}$ can be expressed
in the form of equation (\ref{PL})
\cite{Dolan:2009wf,Spradlin:2009qr}. As discussed in
\cite{Dolan:2009wf}, equation (\ref{PL}) is equivalent to demanding
that the  $k \times (n-k)$ matrix with entries $c^{-1}_{iI}$ has
rank two, which is equivalent to demanding that all $3 \times 3$
sub-determinants of this matrix vanish, giving rise to conditions on
the $c_{iI}$ which are sextic polynomials in the variables. However
even without examining these conditions in detail, it is clear the
conditions are the same swapping the matrix $c_{iI}$ with its
transpose, which is the statement of $G(k,n) = G(n-k,n)$ (i.e.\ parity). Now, under parity, a given $k \times k$ minor $(m_1 m_2
\cdots m_k)$ of $G(k,n)$ is mapped to its complement
$\overline{(m_1 \cdots m_k)}$ in $G(n-k,n)$, where the
$\overline{(\rule{0ex}{1ex}\qquad)}$ denotes that the $(n-k)$ columns that are {\it
not} $m_1, \ldots, m_k$ are used. Explicitly,
\be\label{minor_complement}
\overline{(m_1 \cdots m_k)} = \epsilon_{m_1 \cdots m_k l_1 \cdots l_{n-k}} (l_1 \cdots l_{n-k}).
\ee

Thus, we see that written in
a GL$(k)$-invariant way, the $(k-2)\times(n-k-2)$ Veronese conditions for
some $(n,k)$ are equivalent to the same number of conditions for
$(n,n-k)$ replacing the $k \times k$ minors with their complements.
For instance, consider the case $k=4$, where the Veronese operators
check whether points lie on the degree-3 curve known as the twisted
cubic. (This has been known for a long time---see, e.g. \cite{White:1915}). Any 6 generic points define a twisted cubic. For 7 points, the case with
$k=4$ is the same as $k=3$ that we have already studied: the condition for 7 points to be on a conic can be written
as, e.g., $S_{123456} = 0, S_{123567} = 0$; so to get the condition
for 7 points to lie on a twisted cubic we may just take the parity
conjugate---i.e.\ replace the factor $(1\,2\,3)$ with $\overline{(1\,2\,3)}
= (4\,5\,6\,7)$ and so on. This gives us the pair of conditions for 7
points to lie on the twisted cubic determined by the first 6. But
then we can use this pair of conditions to test that any number of
further points lie on the twisted cubic. In general, for any $k$, any
$k+2$ points like on the degree-$k$ curve, and we can determine the
conditions for $(k+3)$ points to lie on that curve by looking at the
parity conjugate case where $(k+3)$ points must like on a conic.
These are $(k+3 - 5) = (k-2)$ conditions of the form $S_{i_1 \ldots
i_6} = 0$, which we can translate to the original value of $k$ by
replacing $3 \times 3$ minor with its $[(k+3) - 3] \times [(k+3) -
3] = k \times k$ complement. Having determined these $(k-2)$ conditions for
$(k+3)$ particles to lie on the degree-$k$ curve, we get a total of
$(n - (k+3) + 1) \times (k-2) = (k-2)\times(n-k-2)$ conditions for
checking that all $n$ points lie on the curve.

From this discussion, we may conclude that a manifestly GL$(k)$-invariant Grassmannian formulation of the connected
prescription for twistor string theory will necessarily involve a denominator with $(k-2) \times (n-k-2)$ $S$'s, each of which is given as the difference of a product of four minors.

\section{Deformation and Duality}\label{deformation_section}\setcounter{equation}{0}

We have now seen two apparently quite different formulations of Grassmannian theories with a particle interpretation. The first was motivated by unifying the residues of ${\cal L}_{n,k}$ contributing to the tree amplitude into a single algebraic variety, which allowed us to think about adding particles one at a time to construct higher-point amplitudes while keeping the Yangian symmetry manifest. The cyclic invariance of this object is not completely manifest, although at least for NMHV amplitude, the cyclic invariance of the amplitude obtained from $\Gamma^{{\cal L}}$ follows straightforwardly from residue theorems. Finally, the $U(1)$-decoupling identity is not manifest at all.  

One might like to see the cyclic symmetry and $U(1)$-decoupling identities in a much more manifest way. This is what the connected prescription for twistor string theory accomplishes beautifully, by showing that the amplitude is {\it almost} permutation invariant, only breaking down to cyclic invariance because of the ``MHV" factor on the worldsheet $\frac{1}{(\sigma_1 \sigma_2) \cdots (\sigma_n \sigma_1)}$. The price is that  dual superconformal invariance is not manifest. 

Despite appearances, the remarkable statement is that the amplitudes computed in these two apparently very different ways should agree:
\be
{\cal T}_{n,k} = {\cal L}_{n,k}^{\Gamma^{{\cal L}}_{n,k}}.
\ee
We would like to understand why this miracle can happen, beginning with the NMHV amplitudes.  It is a good start that both forms are written as integrals over a single variety---but to go further in making the comparison, we need to deal with the problem that the maps $f_k$ involve the product of {\it three} minors while the Veronese operators involve the product of ${\it four}$ minors. Clearly we need to find a modified form of the $f_k$, which involves a fourth minor. We can also motivate the need for finding a modified form of the $f_k's$ with a fourth minor in another way. Since we will soon be interested in deforming the $f_k$,
in order to have a consistent behavior under the scaling of each column vector of the matrix $C_{\alpha\,a}$---i.e.\ under little group rescalings---we have to deform each component of the map $f_k = (k\smallminus2\,\,k\smallminus1\,\,k)(k\,1\,2)(2\,3\,k\smallminus2)$ by something that preserves the original scaling. Note that it is impossible to {\it add} a polynomial in the minors to $f_k$ to achieve this. However, we can modify each $f_k$ as follows
\be
f^{\rm modif}_k =  (k\smallminus2\,\,k\smallminus1\,\,k)(k\,\,1\,\,2)(2\,\,3\,\,k\smallminus2)(1\,\,3\,\,k\smallminus1).
\ee
By doing this we can deform it while keeping the map holomorphic. The reader might worry about the fact that the new factor $(1\,\,3\,\,k\smallminus1)$ has introduced new poles. It is not hard to show that if $h_n$ is modified as
\be
h_n^{\rm modif} = \frac{\prod_{j=6}^{n-1}[(1\,\, 2\,\, j)(2\,\, 3\,\,j\smallminus1)]\prod_{\ell=5}^{n-1}(1\,3\,\ell)}{(n\smallminus1)(1)(3)}\,\,,
\ee
then the proof presented in section \ref{nmhv_proof_subsection} is not affected.

Even more surprising is the fact that in the new form, $f_k^{\rm modif}$ admits {\it a continuous family of deformations} in such a way that the amplitude is independent of the deformation parameter! Let us denote the deformed $f_k^{\rm modif}$ by $S_k(t_k)$ in anticipation to the connection with the twistor string. More precisely, the deformation we would like to perform is the following
\begin{equation}
\label{eq:comi}
\begin{split}
S_k(t_k) =&\phantom{\,-\,\,\,\,\,\,\,} (k\smallminus2\,\,k\smallminus1\,\,k)(k\,\,1\,\,2)(2\,\,3\,\,k\smallminus2)(k\smallminus1\,\,1\,\,3) \\
& - t_k (k\smallminus1\,\,k\,\,1)(1\,\,2\,\,3)(3\,\,k\smallminus2\,\,k\smallminus1)(k\,\,2\,\,k\smallminus2),
\end{split}
\end{equation}
where $t_k$ is a real parameter (the restriction of reality is to
ensure that for generic $\lambda$'s and $\tilde\lambda$'s, no pole of
the form $1/(i\,\,i\smallplus1\,\,i\smallplus2)$ will be hit by any of the $S_k(t_k)$). (The
minus sign in (\ref{eq:comi}) is introduced for later convenience.)

Let us denote the family of maps $\vS_t\equiv(S_6(t_6),\ldots, S_n(t_n))$. In a moment, we will show that the contour integral
\be
\label{connmhv}
\int\limits_{\vS_t}d^{n-5}\tau \frac{H_n}{S_6(t_6)S_7(t_7)\cdots S_{n}(t_n)}
\ee
is $t$-independent using a contour deformation and global residue theorems. Here, $H_n = h_n^{\rm modif}$. When $t_k =1$, $S_k(1)$ becomes the Veronese operator checking the localization of the
six points $\{k\smallminus2, k\smallminus1, k, 1, 2, 3\}$ on a conic in $\mathbb{CP}^2$, but lacks any convenient geometric interpretation for $t\neq0$.

We have checked by explicitly computing the factor $F(C)$ from
equation (\ref{fofc}), along the lines of the computations in
\cite{Dolan:2009wf,Spradlin:2009qr}, that choosing these Veronese operators to appear in the
holomorphic $\delta$-functions, the numerator factor $H(C)$ precisely
coincides with $h(C)$. Thus, $t$-independence proves the equality
of ${\cal T}_{n,3}$ an ${\cal L}_{n,3}$ equipped with contour
$\Gamma^{{\cal L}}_{n,3}$. As we already remarked, this establishes that the
amplitude satisfies the remarkable $U(1)$-decoupling identity.

It only remains to prove the $t$-independence of the amplitude, which
follows from a straightforward argument using the observations of
\cite{ArkaniHamed:2009sx}. Using the notation of \cite{ArkaniHamed:2009sx}, we think of
one of the $\delta$-function factors as a pole $\frac{1}{d}$, and we use the global residue theorem grouping with the $(n-5)+1$
polynomial factors being the $(n-5)$ $f_i$'s, together with the
remaining three minors in the denominator and $d$, $(n\smallminus1)(1)(3)d$, for
the last polynomial. Now, as in \cite{ArkaniHamed:2009sx}, we deform the pole
away from $d=0$, getting a sum over terms setting $(1)=0, (3)=0$ and
$(n\smallminus1) = 0$. Now, in all of our deformations, the coefficient of $t$
contains a factor $(1\,\,2\,\,3)$, so the term with $(1)=0$ kills the
$t$-dependence of all these terms and is trivially $t$-independent.
The terms with $(n\smallminus1)=0$ and $(3)=0$ make $t$-independent the first and
the last of the $f$'s respectively, and are seen to be $t$-independent
by induction, down to the $n=6$ case which is trivially
seen to be $t$-independent. Note that this argument can also
be thought of as a direct contour-deformation argument relating the
connected prescription of the twistor string theory to the
disconnected prescription given by the CSW rules!

Note that even without this explicit argument, the form of the connected prescription given by equation (\ref{connmhv}) (at $t_k=1$) betrays its connection to CSW. The reason is the presence of the product of three minors $(n\smallminus1)(1)(3)$ in the denominator of $H_n$: the global residue theorem tells us that $\langle (n\smallminus1)(1)(3) \rangle = 0$, where the ``expectation value" is here defined with the integrand of the connected prescription.  But this is a CSW operator! Furthermore, since the twistor string starting point is manifestly cyclically invariant, we must have have that
$\langle (i\smallminus 2) (i) (i\smallplus2) \rangle=0$ for all $i$.  This is a much stronger constraint than the vanishing of the Veronese operators, and is the way the connected prescription alerts us to CSW localization.

For general $k$, we expect a similar analysis to hold. Each of the $f_i$ can be
modified to be written as a product of 4 minors in the form
\be f^{{\rm modif}}_i = M^i_1 M^i_2
M^i_3 M^i_4\,\,.
\ee
We can now consider deformation by a parameter
$t_i$ of the form \be f_i(t) = M^i_1 M^i_2 M^i_3 M^i_4 - t_i M^{\prime
i}_1 M^{\prime i}_2 M^{\prime i}_3 M^{\prime i}_4 \ee and at $t_i=1$, this deformed $f_i$ coincides precisely with
Veronese \mbox{operators $S_i$} \be S_i = M^i_1 M^i_2 M^i_3 M^i_4 -
M^{\prime i}_1 M^{\prime i}_2 M^{\prime i}_3 M^{\prime i}_4\,\,.\ee
Furthermore, for this choice of Veronese operators, the numerator factors in the two forms should become identical
\be
h(C) = H(C).
\ee
In our discussion of N$^2$MHV amplitudes, we will present very strong evidence supporting this claim with direct verification through the 10-point amplitude.
Given this remarkable fact, it is very natural to look for a generalization of the very simple contour deformation argument we gave
for NMHV amplitudes to establish the $t$-independence of the amplitude.

Assuming that the argument holds for all $n$ and $k$, we find not only a duality between ${\cal T}_{n,k}$ and ${\cal L}_{n,k}$ equipped with $\Gamma^{{\cal L}}_{n,k}$,  but equality for an infinite class of theories labeled by the continuous parameter $t$. In a whimsical sense, we might think of $t$ as representing an ``RG" flow. In this analogy the ${\cal L}_{n,k}$ description at $t=0$ is the ``ultraviolet" theory, with the individual residues being the ``gluons", with all symmetries manifest, while the ${\cal T}_{n,k}$ description is the ``infrared" picture with the unified residues combined into ``hadrons", where the ``macroscopic" properties of the collection of residues---the cyclic symmetries and $U(1)$-decoupling identities---are manifest.

\section{NMHV Amplitudes}\label{NMHV_section}\setcounter{equation}{0}

Having described the central ideas of this paper in general terms, we
turn to examining them in detail for the simplest non-trivial case of
NMHV amplitudes. We will begin by showing the sum over residues
with the even/odd/even structure of given by $\Gamma^{{\cal L}}$ in
equation (\ref{nmhvcont}) can be unified into a single variety in a
natural way. We will then show that this ansatz can be $t$-deformed to
the amplitude computed from the connected prescription for twistor
string theory. We end the section by comparing these two ways of
unifying the residues into a single variety.

Let's start by explicitly constructing a holomorphic map
$\vf_n:\mathbb{C}^{n-5}\to \mathbb{C}^{n-5}$ defined in terms of
$n-5$ polynomials $\vf\equiv(f_6,\ldots , f_{n})$ and a function $h_n$, such the
tree level amplitude is given as
\be
A^{(3)}_{n} =\int\limits_{\vf_n=0} d^{n-5}\tau
\frac{h_n}{f_6\cdot f_7\cdots f_{n}}\,.
\ee
The reason for the offset in the labeling of the polynomials $f_i$ will become clear below. The construction is such that taken as rational functions one has,
\be
\label{eq:expi}
\frac{h_n}{f_6\cdot f_7\cdots f_{n}} = \frac{1}{(1\,2\,3)(2\,3\,4)\cdots (n\,1\,2)}.
\ee
It is natural to try to construct the map $f$ from consecutive
minors as those are the ones that enter in (\ref{eq:expi}). However,
it is easy to see that for $n\geq 8$ it is impossible to construct a
holomorphic map from consecutive minors such that the contour given
in \cite{ArkaniHamed:2009dn} is contained in the set of zeros of
the map. It is instructive to see the obstruction already for $n=8$.
The contour $\Gamma_{8,3}^{{\cal L}}$ is given by
\begin{equation}
\label{eq:ACCK8}
\begin{split}\Gamma_{8,3}^{{\cal L}} = &\phantom{\,+\,\,} (1)(2)\left[(3)+(5)+(7)\right]+
 (3)(4)\left[(5)+(7)\right]+(5)(6)(7)
 \\ &+(1)(4)\left[(5)+(7)\right]\,\,\, \qquad+ (3)(6)(7)\\
&+(1)(6)(7).
\end{split}
\end{equation}

Let's try to construct a mapping $\vf_8:\mathbb{C}^3\to \mathbb{C}^3$, with $f_i$ polynomials in the minors $(k)$.
Consider the terms $(1)(2)(3)$, $(1)(4)(5)$ and $(3)(4)(5)$. From the
first term we learn that $(1)$ and $(3)$ must belong to different
$f_i$'s, while combining the information from the second and third we
learn that $(1)$ and $(3)$ must be on the same $f_i$, which is a
contradiction.

Having seen the need for a different way to construct $\vf_n$ we now
show that the construction is very natural and recursive. The reason
it is recursive has a beautiful physical interpretation: it is equivalent
to the operation of adding one particle at a time!

In order to motivate the construction, consider first the
six-particle amplitude. (In this section, $k$ is always $3$ and will
therefore be frequently suppressed). The contour given in \cite{ArkaniHamed:2009dn} is
\linebreak\mbox{$\Gamma_{6,3}^{{\cal L}}= (2\,3\,4)+(4\,5\,6)+(6\,1\,2)$}. By this we mean three
terms, the first of which is
\be
\int\limits_{(2\,3\,4)=0} d\tau \frac{1}{(1\,2\,3)(2\,3\,4)(3\,4\,5)(4\,5\,6)(5\,6\,1)(6\,1\,2)}.
\ee
Clearly, if we define the map $f_6:\mathbb{C}\to \mathbb{C}$ as
$f_6=(2\,3\,4)(4\,5\,6)(6\,1\,2)$, then 
\be
A_{6}^{(3)}=\int\limits_{f_6=0} \!\!\!d\tau\,\, \frac{h_6(\tau)}{f_6(\tau)}
\ee
with $h_6 = 1/(1\,2\,3)(3\,4\,5)(5\,6\,1)$.

In order to find a recursive way of constructing the map for all $n$,
let us consider the five particle integrand,
\be
\label{eq:five}
\frac{1}{(1\,2\,3)(2\,3\,4)(3\,4\,5)(4\,5\,1)(5\,1\,2)},
\ee
and ask what factor would convert this into the six-particle integrand. Clearly,
\be
\inversesoft{5\to6}{k=3} = \frac{1}{(5\,6\,1)}\times\frac{(4\,5\,1)(5\,1\,2)(2\,3\,4)}{f_{6}},
\ee
where $f_{6}=(4\,5\,6)(6\,1\,2)(2\,3\,4)$, does what is needed. It might be
puzzling at first why we introduced $(2\,3\,4)$ both in the numerator
and in the denominator. The reason for this is clear from the previous
discussion. Recall that we have to define $h_6$ and $f_6$
independently. Multiplying (\ref{eq:five}) by $\inversesoft{5\to6}{}$ we immediately
find $h_6$.

We interpret the operation of multiplying by $\inversesoft{5\to6}{}$ as that of adding
particle six to the five-particle amplitude. We will see that this
interpretation is justified when we show that in general this
corresponds to building an object with the right holomorphic soft-limit.

\subsection{Recursive Construction}

From the six-particle example, we are motivated to construct the
$n$-particle amplitude recursively as follows. Let
$\vf_{(n\text{-}1)}:\mathbb{C}^{n-6}\to \mathbb{C}^{n-6}$ be the holomorphic
map and $h_{n-1}$ the meromorphic function such that
\be
A^{(3)}_n = \int\limits_{\vf_{(n\text{-}1)}=0} d^{n-6}\tau
\frac{h_{n-6}}{f_6\,f_7\cdots f_{n-1}}.
\ee
Then the $n$-particle amplitude is obtained by ``multiplying" the integrand by
\be
\inversesoft{(n-1)\to n}{} = \frac{1}{(n\smallminus1\,\,n\,\,1)}\times\frac{(n\smallminus2\,\,n\smallminus1\,\,1)(n\smallminus1\,\,1\,2)(2\,3\,n\smallminus2)}{f_{n}}
\ee
with $f_{n}=(n\smallminus2\,\,n\smallminus1\,\,n)(n\,\,1\,\,2)(2\,\,3\,\,n\smallminus2)$. By ``multiplying"
we mean extending the map $(f_6,f_7,\ldots ,f_{n-1})$ to a map
$\vf_{n}:\mathbb{C}^{n-5}\to \mathbb{C}^{n-5}$ by adding $f_{n}$ as
the last component---i.e.\ , forming $\vf_{n}=(f_6,f_7,\ldots ,f_{n-1},f_{n})$.
Likewise, we have a new $h_n$ given by
\be
h_n = h_{n\smallminus1}\frac{(n\smallminus2\,\,n\smallminus1\,\,1)(n\smallminus1\,\,1\,\,2)(2\,\,3\,\,n\smallminus2)}{(n\smallminus1\,\,n\,\,1)}.
\ee
Note that what we are doing can be interpreted as adding the particle
$n$ between \mbox{$(n-1)$ and $1$}:\\
\[\includegraphics[scale=0.85]{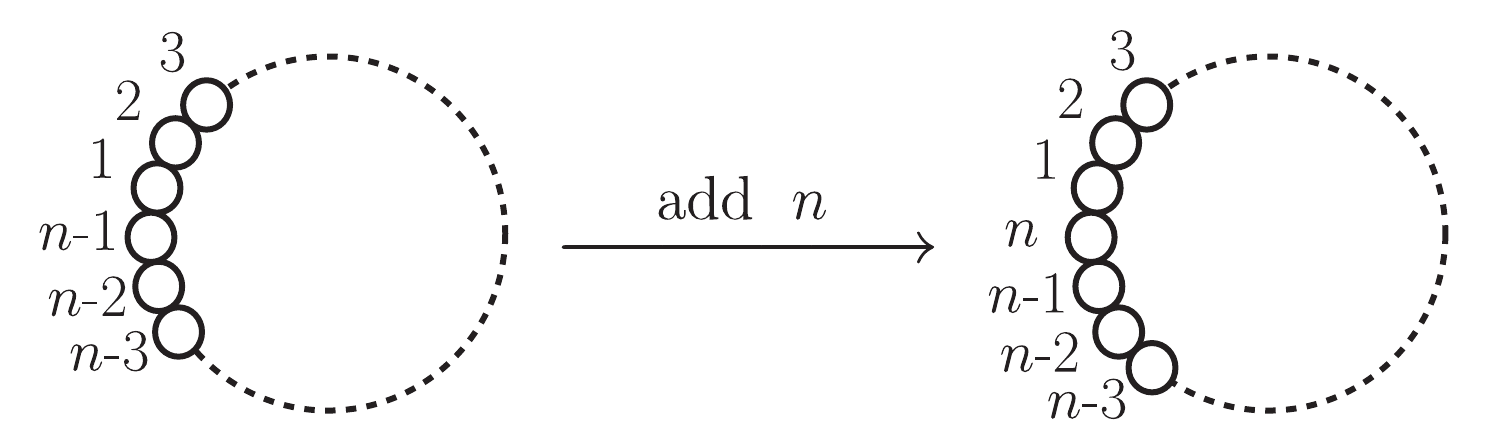}\nonumber
\]

Given that we are dealing with $3 \times 3$ minors for NMHV
amplitudes, it is reasonable that the ``add particle $n$" operation
could involve particles $(n-3)$ up to $3$. There are a number of
choices we could make for how to do this, but the one we have
presented accomplishes the task of unifying the residues in the nicest
way that also manifests a number of important properties that we will
discuss at greater length at the end of this section.

\subsection{The $n=7,8$ Amplitudes}

For now, let us show how this construction works explicitly for $n=7$
and $n=8$. The seven
particle NMHV contour is given by
\be
\label{eq:ampi}
\Gamma_{7,3}^{{\cal L}} = (2)\left[ (3)+(5)+(7)\right] + (4)\left[(5)+(7)\right] + (6)(7).
\ee
Using the recursive construction, we multiply the six-particle $h_6/f_6$ by
\be
\inversesoft{6\to7}{}  = \frac{1}{(6\,7\,1)}\times\frac{(5\,6\,1)(6\,1\,2)(2\,3\,5)}{f_{7}}
\ee
with $f_7 = (5\,6\,7)(7\,1\,2)(2\,3\,5)$.

Putting everything together we find the seven-particle amplitude to be
\be
A_7^{(3)}=\int\limits_{\vf_7=0} d^2\tau \frac{h_7(\tau)}{f_6(\tau)f_7(\tau)} \quad {\rm
with}\quad h_7(\tau)
=\frac{(6\,1\,2)(2\,3\,5)}{(6\,7\,1)(1\,2\,3)(3\,4\,5)},
\ee
while the map $\vf_7=(f_6,f_7)$ where,
\be\label{eq:kai}
f_6(\tau) = (2\,3\,4)(4\,5\,6)(6\,1\,2)\quad\mathrm{and}\quad f_7(\tau) = (5\,6\,7)(7\,1\,2)(2\,3\,5).
\ee
The claim is that the tree-level contour is nothing but the sum over
the residues of all the $9$ zeros of $\vf_7$. At first sight this might seem
surprising because by na\"{i}vely simplifying $h_7/(f_6f_7)$ one would
find the original object
\be
\frac{1}{(1\,2\,3)(2\,3\,4)(3\,4\,5)(4\,5\,6)(5\,6\,7)(6\,7\,1)(7\,1\,2)},
\ee
integrated over $[(2)+(4)][(5)+(7)]$. This only gives four terms of
the six terms in (\ref{eq:ampi}) and therefore it cannot be the
correct amplitude. The resolution to this na\"{i}ve puzzle is that we
should not cancel terms and forget about them! Recall that the map $\vf_7$ is
independent of the function $h$ and we are supposed to carefully study
all $9$ residues. It turns out that only six are nonzero, and these add up
to the amplitude. Among the six, four of them are the ones we got from
the na\"{i}ve analysis. Let us present the other two.

The first term missed in the na\"{i}ve cancelation is the residue at
the point located where $(2\,3\,4)=0$ and $(2\,3\,5)=0$. Note that $(2\,3\,5)$ is
also a factor in the numerator, and this is why na\"{i}vely may not be expected to contribute. The reason it does contribute is that when we impose the condition that the points $2,3,5$ be (projectively) collinear {\it and} points
$2,3,4$ be collinear, it follows that $3,4,5$ must also be
collinear, and hence $(3\,4\,5)=0$. But $(3\,4\,5)$ is a factor in the
denominator of $h_7$ and therefore is a pole with non-vanishing residue. In order to compute
the residue in these cases we will use the following simple result:
given linear polynomials, $A,B$ and $C$ in two variables, such that
$C=0$ when $A=B=0$ one has the identity
\be
\int\limits_{|A|=\epsilon_1, |B|=\epsilon_2}\hspace{-0.5cm} d^2\tau \frac{A}{ABC} =
\int\limits_{|B|=\epsilon_1, |C|=\epsilon_2}\hspace{-0.5cm} d^2\tau \frac{A}{ABC} =
\int\limits_{|B|=\epsilon_1, |C|=\epsilon_2}\hspace{-0.5cm} d^2\tau \frac{1}{BC}\,\,,
\ee
for any $\epsilon_1$ and $\epsilon_2$ arbitrarily small. This means
that what we called the residue at $A=B=0$ is the same as the residue
at $B=C=0$.

Using the identity we find that the pole at $(2\,3\,4)=(2\,3\,5)=0$ can also
be thought of as a pole at $(2\,3\,4)=(3\,4\,5)=0$. Canceling $(2\,3\,5)$ in the
numerator and the denominator we find that it is what we call residue
$(2)(3)$.

The second term is at $(6\,1\,2)=(7\,1\,2)=0$. At this point we also have
$(6\,7\,1)=0$ which is a pole of $h_7$. Using the same identity one finds
the residue $(6)(7)$.

All other remaining 3 out of the original 9 residues vanish due to the factors in the
numerator as they do not set any other factors in the poles $h_7$ to zero.

Putting together the first four terms we found in the na\"{i}ve analysis
plus the two new terms we find (\ref{eq:ampi})
\be
(2)\left[ (3)+(5)+(7)\right] + (4)\left[(5)+(7)\right] + (6)(7).
\ee

\subsubsection{Aside: A Subtlety in the Use of the Global Residue Theorem}\label{GRT_subtlety_section}

Before continuing on to the eight particle example, it is important to
discuss a subtlety which appears in the application of the global
residue theorem (GRT) to
residue integrals of the sort we are dealing with. In fact, as we will
illustrate for the seven
particle example, a na\"{i}ve application of the global residue theorem
leads to a contradiction.
Let us recall that the global residue theorem asserts that given a
holomorphic map $f:\mathbb{C}^m\to \mathbb{C}^n$ with $m\leq n$ and a
holomorphic function $s$ in $\mathbb{C}^m$, then for any way of
constructing a map $g:\mathbb{C}^m\to \mathbb{C}^m$ by combining
several $f_i$'s into single $g_i$'s such that $g$ only has isolated
zeros then
\be
\label{eq:GRT}
\sum_{p\in g^{-1}(0)} \int_{T^m_p} d^m\tau
\frac{s(\tau)}{f_1(\tau)\cdots f_m(\tau)f_{m+1}(\tau)\cdots f_n(\tau)}
= 0
\ee
where the sum is over all zeros of $g$ and the contour $T^m_p$ is
defined by translating $p\in \mathbb{C}^m$ to the origin and having
$|g_i|=\epsilon_i$ with $\epsilon_i$ a sufficiently small positive
real number. The theorem holds provided there is no contribution at
infinity, which is true when \linebreak \mbox{${\rm deg}\, s\leq \sum_{i=1}^m {\rm
deg}\, g_i - (m+1)$}.
Suppose that the $i^{\rm th}$ component of $g$ is given by $g_i = f_k
f_l$ for some $k$ and $l$. Using (\ref{eq:GRT}) one could conclude
that
\be
\label{eq:nai}
\sum_{p\in \Gamma_k} \int_{T^m_p} d^m\tau
\frac{s(\tau)}{f_1(\tau)\cdots  f_n(\tau)} = - \sum_{p\in \Gamma_l}
\int_{T^m_p} d^m\tau \frac{s(\tau)}{f_1(\tau)\cdots f_n(\tau)},
\ee
where $\Gamma_k$ (or by $\Gamma_l$) are the zeros of the map $g$
where $g_i$ is replaced by $f_k$ (or by $f_l$). In one complex
dimension this is the usual way Cauchy's theorem is applied.
Consider now the 7-particle amplitude. We can set $m=2$, $s(\tau)
= (6\,1\,2)(2\,3\,5)$, and introduce $f_5=(6\,7\,1)(1\,2\,3)(3\,4\,5)$ in addition to
$f_6$ and $f_7$. This gives a map $f^{\rm new}: \mathbb{C}^2\to
\mathbb{C}^3$. According to the theorem we have to construct a map
$g:\mathbb{C}^2\to \mathbb{C}^2$ out of the three components of
$f^{\rm new}$. One possible choice is $g_1=f_6$ and $g_2=f_5f_7=
(6\,7\,1)(1\,2\,3)(3\,4\,5)\, f_7$, with $f_6$ and $f_7$ given in (\ref{eq:kai}).
Recalling that each minor is linear in $\tau$'s we find that the
degree condition for the application of the GRT is satisfied.
Using (\ref{eq:nai}) one finds
\be
\int\limits_{\{f_6, f_7\}} \!\!d^2\tau \,\,\frac{(6\,1\,2)(2\,3\,5)}{(6\,7\,1)(1\,2\,3)(3\,4\,5)\,\, f_6\,f_7} =
- \int\limits_{\{f_5,f_6\}} \!\!d^2\tau \,\,\frac{(6\,1\,2)}{(5\,6\,7)(7\,1\,2)\,\, f_5\, f_6}\,\,.
\ee
The LHS has been shown to give $A^{(3)}_7$ in the first part of
this section. Let us now compute the RHS where the contour is a sum
over the zeros of $\{(6\,7\,1)(1\,2\,3)(3\,4\,5),f_6\}$. A straightforward
computation reveals that this is the sum over the usual residues of
${\cal L}_{n,k}$ given by
\be
\label{eq:bad}
-(6)[(4)+(2)+(7)]-(1)[(4)+(2)]-(3)[(4)+(2)].
\ee
We can use a GRT as was done in
\cite{ArkaniHamed:2009dn} to bring this into a more recognizable form. We will use that
$(6)[(1)+(2)+(3)+(4)+(5)+(7)]=0$ in (\ref{eq:bad}) and a rearrangement
of terms (recalling that $(i)(j)=-(j)(i)$) to get
\be
\label{eq:uhh}
-(1)[(2)+(4)+(6)]-(3)[(4)+(6)]-(5)(6) + (2)(3).
\ee
The first six terms give rise to the parity-conjugate version of the BCFW-contour as explained in \cite{ArkaniHamed:2009dn} and therefore equal $A^{(3)}_7$.
This means that (\ref{eq:uhh}) equals
\be
A^{(3)}_7 + (2)(3),
\ee
which is a contradiction, as advertised.
As mentioned at the beginning of the discussion, there is an implicit
assumption in using the GRT (\ref{eq:GRT}) to derive (\ref{eq:nai}).
The implicit assumption is that $\Gamma_k$ and $\Gamma_l$ as sets of
points in $\mathbb{C}^m$ are disjoint. This is exactly what fails in
our seven particle example. Indeed, note that the point $(2)=(3)=0$
appears in both contours! In order to see this note that the map
defined by $g_1=f_6$ and $g_2=(6\,7\,1)(1\,2\,3)(3\,4\,5)\, f_7$, with
$f_7=(5\,6\,7)(7\,1\,2)(2\,3\,5)$, has a double zero at $(2\,3\,4)=(3\,4\,5)=0$ since
$(2\,3\,5)$ also vanishes there.
This means that while the GRT is valid as given in (\ref{eq:GRT}), the
splitting into two parts must be defined independently in this
situation. In other words, one has to decide where to keep $(2)(3)$.
In our construction we have defined the amplitude in such a way that
$(2)(3)$ is kept where the contour is defined by $\{ f_6,f_7\}$ and
therefore should subtracted from the second form, i.e.\ ,
\be
A^{(3)}_7 = - \int\limits_{\{f_5,f_6\}} \!\!\!d^2\tau\,\,\,
\frac{(6\,1\,2)}{(5\,6\,7)(7\,1\,2)\,\,f_5\,f_6} - (2)(3)\,\,.
\ee
This is very reminiscent of what happened in \cite{Dolan:2009wf}, where
some forms for the connected prescription gave rise to the amplitude
only after subtracting ``spurious" configurations. 
Note that the same exercise can be repeated but using $g_1=f_5f_6$ and
$g_2=f_7$. We leave it to the reader to show that the same phenomena
happens when this time the shared point is given by $(6)=(7)=0$. Recall
that $(2)(3)$ and $(6)(7)$ were precisely the special points in the
previous discussion of the seven particle amplitude.

\subsubsection{Eight-Particle Example}

The eight particle amplitude can be analyzed in a similar manner to the seven particle example.
Following the same steps as before we find
\be
\int\limits_{\vf_8=0} \!d^3\tau\,\, \frac{h_8(\tau)}{f_6f_7f_8} \quad {\rm with}\quad
h_8(\tau) =\frac{(6\,1\,2)(2\,3\,5)(7\,1\,2)(2\,3\,6)}{(7\,8\,1)(1\,2\,3)(3\,4\,5)}
\ee
while the map $\vf_8\equiv(f_6,f_7,f_8)$ and for which the $f_i$ are given by
\be
f_6 = (2\,3\,4)(4\,5\,6)(6\,1\,2), \quad f_7 = (5\,6\,7)(7\,1\,2)(2\,3\,5),\quad
f_8= (6\,7\,8)(8\,1\,2)(2\,3\,6).
\ee
Once again, the na\"{i}ve cancelation of terms when $h_8/(f_6f_7f_8)$ is
thought of as a rational function leads the contour
$[(2)+(4)](5)[(6)+(8)]$ which is clearly wrong as it misses 6 terms!

Four of the missing terms are of the same origin as the two missing
terms in the seven particle amplitude. We simply list the map and
leave the geometric proofs an elementary exercises for the reader:
\begin{equation}
\begin{split}
\{(2\,3\,4),(2\,3\,5),(6\,7\,8)\}  \longrightarrow &\, \{(2\,3\,4),(3\,4\,5),(6\,7\,8)\} =
(2)(3)(6); \\
\{(2\,3\,4),(2\,3\,5),(8\,1\,2)\}  \longrightarrow &\, \{(2\,3\,4),(3\,4\,5),(8\,1\,2)\} =
(2)(3)(8); \\
\{(2\,3\,4),(7\,1\,2),(8\,1\,2)\}  \longrightarrow &\, \{ (2\,3\,4),(7\,8\,1),(8\,1\,2)\} =
(2)(7)(8); \\
\{(4\,5\,6),(7\,1\,2),(8\,1\,2)\}  \longrightarrow &\, \{ (4\,5\,6),(7\,8\,1),(8\,1\,2)\} = (4)(7)(8).
\end{split}
\end{equation}
The final two missing terms are more interesting. One of the missing
terms from the $\mathcal{L}_{n,k}$-contour is $(2)(3)(4) = \{(2\,3\,4),(3\,4\,5),(5\,6\,7)\}$.
Note that this singularity has the geometric interpretation of
imposing that points $2,3,4,5,6$ and $7$ be collinear in the
$\mathbb{CP}^2$-sense.

Let us now look at the map $\vf_8$ at the point $(2\,3\,4)=(2\,3\,5)=(2\,3\,6)=0$.
Note that this imposes exactly the same geometric constraint and it is
therefore the same point in $(\tau_1,\tau_2,\tau_3)$ space. Since by
construction we have zeros in $h_8$ where $(2\,3\,5)=0$ and $(2\,3\,6)=0$ we
need two poles in the denominator to vanish. These are $(4\,5\,6)$ in
$f_6$ and $(3\,4\,5)$ in $h_8$. Recalling that the residue is computed
using a $T^3$-contour $|(2\,3\,4)|=\epsilon_1$, $|(2\,3\,5)|=\epsilon_2$ and
$|(2\,3\,6)|=\epsilon_3$ one can show that the answer is the same as if we
used the contour $|(2\,3\,4)|=\epsilon_1$, $|(3\,4\,5)|=\epsilon_2$ and
$|(4\,5\,6)|=\epsilon_3$ and therefore the residue is identical to what we
call $(2)(4)(5)$.

Moreover, this also shows that the same point in $\mathbb{C}^3$ is determine by
$(4\,5\,6)=(2\,3\,5)=(2\,3\,6)=0$. This means that this is not a distinct zero of $\vf_8$ and
therefore does not give rise to a new residue.

Exactly the same happens to the second missing term but this time we
have to start with $\{(6\,1\,2),(7\,1\,2),(8\,1\,2)\}$ and realize that $(6\,7\,8)$ in
$f_8$ and $(7\,8\,1)$ in $h_8$ vanish. Summarizing the new kind of terms
\begin{eqnarray}
\{(2\,3\,4),(2\,3\,5),(2\,3\,6)\} = \{ (4\,5\,6),(2\,3\,5),(2\,3\,6)\} & \longrightarrow & \{(2\,3\,4),(3\,4\,5),(4\,5\,6)\} =
(2)(3)(4);  \nonumber \\
\{(6\,1\,2),(7\,1\,2),(8\,1\,2)\} = \{ (6\,1\,2),(7\,1\,2),(6\,7\,8)\} & \longrightarrow & \{ (6\,7\,8),(7\,8\,1),(8\,1\,2)\} = (6)(7)(8); \nonumber
\end{eqnarray}
and collecting all these results we find 10 residues which agree with
$\Gamma^{{\cal L}}_{8,3}$ given in (\ref{eq:ACCK8}).

\newpage
\subsection{General Proof For All $n$}\label{nmhv_proof_subsection}

Let us now prove that
\be
A_{n}^{(3)}=\int\limits_{\vf_n}\!\!\frac{h_n}{f_6\,f_7\cdots f_n},
\ee
reproduces the correct tree-level amplitude as defined by
$\Gamma^{{\cal L}}_{n,3}$ for all NMHV amplitudes in full generality. The
proof proceeds by induction. In fact, it is a simple generalization of
the computation we have already seen for eight particles---which is the
simplest case where all the general ingredients appear.

Let us state more precisely what we want to prove. Consider the
$n$-particle amplitude. Given that as rational functions
\be
\label{eq:vitu}
\frac{1}{(1)(2)\cdots (n\smallminus2)(n\smallminus1)(n)} = \frac{h_{n}}{f_6\cdot f_7
\cdots f_{n-1}\cdot f_{n}}\,\,,
\ee
all we need to show is that the points in $\mathbb{C}^{n-5}$ determined by
\be
\label{eq:poco}
\begin{split}& \underbrace{{\cal E}_{n} \star {\cal O}_{n} \star {\cal E}_{n} \star
\cdots}\\ & (n-5) \, {\rm factors}
\end{split}\ee
are zeros of $\vf_n$. These zeros are guaranteed to give
the right residues while all other zeros of $\vf_n$ have
zero residue by virtue of (\ref{eq:vitu})! Recall from
\cite{ArkaniHamed:2009dn} that the $\star$-product is such that
$(i)\star (j)=0$ if $i>j$, and \be {\cal E}_n = (2)+(4)+\ldots + (2[n/2])\quad\mathrm{and}\quad
{\cal O}_{n}=(1)+(3)+\ldots + (2[n/2]+1).\ee
A note on notation: in this discussion we use $(i)$ for a consecutive
minor of the $n$-particle amplitude. Any other minor will be written
explicitly as $(i\,\,j\,\,k)$.

\subsubsection{Induction Argument}

Start by assuming that the statement is true for $(n-1)$-particles. In
other words, we can freely start with
\be\label{eq:kiko}
\frac{1}{(1)(2)(3)\cdots (n\smallminus3)(n\smallminus2\,\,n\smallminus1\,\,1)(n\smallminus1\,\,1\,\,2)}
\ee
and consider only the zeros of $f_{(n-1)}$ corresponding to
\be\begin{split}
& \underbrace{{\cal E}_{n-1} \star {\cal O}_{n-1} \star {\cal E}_{n-1}
\star \cdots}\\ & (n-6) \, {\rm factors}
\end{split}
\ee
where the subscript is there to indicate that the minors in
(\ref{eq:kiko}) are being used.

Recall that in order to get the $n$-particle formula all we have to do
is to multiply by $h_{n-1}/f_6\cdots f_{n-1}$ by
\be
\inversesoft{(n-1)\to n}{(3)} = \frac{(n\smallminus2\,\,n\smallminus1\,\,1)(n\smallminus1\,\,1\,\,2)(n\smallminus2\,\,2\,\,3)}{(n\smallminus1\,\,n\,\,1)\,\, f_{n}}
\ee
with $f_{n} = (n\smallminus2)(n)(n\smallminus2\,\,2\,\,3)$. For the purpose of the
proof, all we need to show is that all the points in $\mathbb{C}^{n-5}$
given by (\ref{eq:poco}) are also points in
\be
\label{eq:sapa}
\left[ {\cal E}_{n-1} \star {\cal O}_{n-1} \star {\cal E}_{n-1} \star
\ldots \right] \times \left[ (n\smallminus2)+(n)+(n\smallminus2\,\,2\,\,3)\right].
\ee
The multiplication sign `$\times$' is there to stress that every single
term on the left must be multiplied by every term on the right (unlike
the symbol $\star$).

The first two terms in the last factor of (\ref{eq:sapa}), i.e.\ , $[(n-2)]$ and $[(n)]$, directly
give terms in (\ref{eq:poco}) except when they hit terms of the form
$[\ldots \star (n\smallminus1\,\,1\,\,2)]$ or $[\ldots \star (n\smallminus2\,\,n\smallminus1\,\,1)\star
(n\smallminus1\,\,1\,\,2)]$. The reason for splitting these two cases will become
clear in a moment.

Terms of the form $[\ldots \star (n\smallminus1\,\,1\,\,2)]\times (n\smallminus2)$ vanish because
no other consecutive minor is set to zero, while terms of the form
$[\ldots \star (n\smallminus1\,\,1\,\,2)]\times (n)$ make $(n\smallminus1\,\,n\,\,1)=0$ and give rise
to $[\ldots \star (n\smallminus1)](n) =[\ldots] \star (n\smallminus1)\star (n) $.
The situation is different and much more interesting for the second
class. Note that $[\ldots]\star (n\smallminus2\,\,n\smallminus1\,\,1)\star (n\smallminus1\,\,1\,\,2)\times (n\smallminus2)$
and $[\ldots] \star (n\smallminus2\,\,n\smallminus1\,\,1)\star (n\smallminus1\,\,1\,\,2)\times (n)$ define the
same point in $\mathbb{C}^{n-5}$! This particular point is precisely
the one where minors $(n\smallminus2)=(n\smallminus1)=(n)=0$. This means that they give
rise to the terms in (\ref{eq:poco}) of the form $[\ldots]\star
(n\smallminus2)\star (n\smallminus1)\star (n)$.

This shows that as sets of points in $\mathbb{C}^{n-5}$
\be
\left[ {\cal E}_{n-1} \star {\cal O}_{n-1} \star {\cal E}_{n-1} \star
\cdots\right]\star \left[ (n\smallminus2) + (n) \right] = \left[ {\cal E}_{n}
\star {\cal O}_{n} \star {\cal E}_{n} \star \cdots\right]\star \left[
(n\smallminus2) + (n) \right]
\ee
The only difference between this formula and what we want is a $(n\smallminus4)$
term in the final factor. The reason is that with $(n-5)$ total factors,
the $\star$-product forces any factor of the form $(n\smallminus k)$ with $k\geq
2$ in the last factor to vanish in (\ref{eq:poco}). Moreover, it is
clear that only one term in (\ref{eq:poco}) has $(n\smallminus4)$ as the final
factor. This is the term \nopagebreak$(2)\star (3)\star (4)\star \ldots \star
(n\smallminus5)\star (n\smallminus4)$. In order to generate this term note that
$(n\smallminus2\,\,2\,\,3)=0$ in (\ref{eq:sapa}) together with \mbox{$(2)=(3)=\ldots
=(n\smallminus1)=0$} implies that $(n\smallminus4)$, which explicitly is given by
$(n\smallminus4\,\,n\smallminus3\,\,n\smallminus2)$, vanishes which is what we wanted to show.

As an aside, note that this proof motivates us to write the ${\cal L}_{n,k}$-contour
as $\star$-multiplication of the $(n-1)$-particle contour by
$[(n)+(n\smallminus2)+(n\smallminus4)]$, in other words, it shows that it is given as
\be
\label{eq:newcon}
[(6)+(4)+(2)]\star [(7)+(5)+(3)]\star [(8)+(6)+(4)]\star \cdots \star
[(n)+(n\smallminus2)+(n\smallminus4)].
\ee

Note that we have unified the residues of this contour into a single
variety; both the contour itself as well as the unification are not
manifestly cyclically invariant. The cyclic invariance of
$\Gamma_{n,3}^{{\cal L}}$ was shown to follow simply from the global residue
theorem in \cite{ArkaniHamed:2009dn}, and hence the unified form we have given it
also gives rise to a cyclically invariant amplitude.

\subsection{``Inverse-Soft" Interpretation}
It remains to show that the ``add one particle at a time" construction
we have given has an interpretation more specifically as an
``inverse-soft" operation, by showing that the multiplicative factor
$\!\inversesoft{(n-1)\to n}{(3)}$ turns into the soft factor for particle $n$ in
the limit $\lambda_n \to 0$. Recall that
\be
\inversesoft{(n-1)\to n}{(3)} =
\frac{(n\smallminus2\,\,n\smallminus1\,\,1)(n\smallminus1\,\,1\,\,2)(n\smallminus2\,\,2\,\,3)}{(n\smallminus1\,\,n\,\,1)\,\, f_{n}}
\ee
with
\be
f_{n} = (n\smallminus2)(n)(n\smallminus2\,\,2\,\,3).
\ee
Now, in order to exhibit the soft limit, we will use the global
residue theorem, choosing $(n-6)$ of the polynomials to be the $f$'s for
the $(n-1)$-particle amplitude, and the remaining polynomial to be $f_n$ times the remaining denominator factors, which among others include
the minor $(n\smallminus1\,\,n\,\,1)$. The residue theorem gives us a sum over terms
putting the remaining denominator factors to zero. It is easy to show
in general (as will be discussed in detail in \cite{ABCCK:2010}), that
none of these contributions can be singular in the soft limit, except
the one where the minor $(n\smallminus1\,\,n\,\,1)$ is set to zero. Focusing only on
this contribution, it will also be shown that {\it every} residue of
${\cal L}_{n,3}$ setting $(n\smallminus1\,\,n\,\,1)$ and any other collection of
minors to zero maps, in the soft limit $\lambda_n \to 0$, to the usual
soft factor multiplied by the corresponding residue of $G(3,n-1)$
determined by the vanishing of these other minors. This guarantees
that the soft limits are manifest as claimed.

\pagebreak\subsection{Connection to the Twistor String}\label{twistor_string_subsection}

As already mentioned in \mbox{section \ref{deformation_section}}, there is a continuous deformation of the map $f_{(n)}$ which does not affect the sum over residues and which gives rise to an integral over the Grassmannian which can be shown to come from the twistor string formulation of the amplitude and which wonderfully manifests the cyclic-symmetry and $U(1)$-decoupling identities of the amplitude.

It is instructive to note that both the cyclic invariance and $U(1)$-decoupling identities can be established without performing the
explicit calculation relating our form of the object to the connected
prescription. By construction, the Veronese operators localize
the integral over the $C_{\alpha\,a}$'s to be over matrices with the Veronese form;
computing the residue tells us to look at what is happening to first
order in a Laurent expansion in $(n-5)$ variables in the vicinity of
the Veronese form. Let us consider such a first-order perturbation
away from the Veronese form given by the following parametrization of
the $C_{\alpha\,a}$ matrix,
\be
C =
\left(\begin{array}{cccc} \xi_1+\sum_{j=1}^{n-5}\epsilon_j \rho^j_1 &
\xi_2+\sum_{j=1}^{n-5}\epsilon_j \rho^j_2 & \ldots & \xi_n
+\sum_{j=1}^{n-5}\epsilon_j \rho^j_n  \\ \xi_1\rho_1  & \xi_2\rho_2 &
\ldots &  \xi_n\rho_n \\ \xi_1\rho_1^2  & \xi_2\rho_2^2 & \ldots &
\xi_n\rho_n^2 \end{array} \right),
\ee
one finds that the leading order in $\epsilon$ of the Veronese
polynomials is linear in $\epsilon$ and can be denoted by $S_k^{\rm
leading}(1)$. This means that the following change of variables
$u_k=S_k^{\rm leading}(1)$ from $(\epsilon_1,\ldots, \epsilon_{n-5})$
to $u_k$ is linear and the contour integral around the point $S_k^{\rm
leading}=0$ can be written as follows
\be
G(\xi_i, \rho_i) = \int \!d^{n-5}u\,\, \frac{1}{u_6 u_7\cdots u_n}\,\,,
\ee
where the contour computes the residue at $u_k=0$ which gives one. Of
course, to get the final result for the tree amplitude one would still
have to integrate over the $\rho$'s, but this form already allows us
to see both the cyclic-symmetries and $U(1)$-decoupling identity. This
is because straightforward computation of the function
$G(\xi_i,\rho_i)$ reveals a very beautiful property: it is almost
permutation invariant. In fact, it is given by
\be
G(\xi_i,\rho_i) = \frac{1}{(\rho_1-\rho_2)(\rho_2-\rho_3)\cdots
(\rho_n-\rho_1)}\times {\tilde G}(\xi_i,\rho_i)
\ee
where ${\tilde G}(\xi_i,\rho_i)$ is fully permutation invariant!
Despite the non-manifest cyclic invariance of this integrand, this
residue {\it is} cyclically invariant, and this conclusion is not
changed in performing the integral over $\rho$'s giving the tree
amplitude. Similarly, since the only breaking of permutation
invariance is in the pre-factor, which is just the same twistor-string
measure guaranteeing the $U(1)$-decoupling identity.

\section{Generalization to N$^2$MHV}\label{NNMHV_section}\setcounter{equation}{0}
Returning to the Grassmannian, it is not difficult to extend our results for general NMHV amplitudes to higher-$k$ by first using parity-conjugation to obtain the contour for $\overline{\text{NMHV}}$, and then view this as the result of having added a particle to an $\overline{\text{MHV}}$ amplitude. It will be instructive to work this out in detail for N$^{2}$MHV, because there are several new structures that emerge first for $k=4$ that will be important for all higher-$k$; these new structures will be discussed in section \ref{residue_geometry}. After deriving a general formula (\ref{generaln2mhv}) for the N$^2$MHV amplitude computed in the Grassmannian, we will check it in detail for the 8-particle amplitude in \mbox{section \ref{8pointsection}}. This will allow us to discuss many of the new structures that emerge beyond NMHV, and which are prerequisite to understanding higher-$k$.

The method by which we will obtain the contour for N$^2$MHV is roughly as follows. We will first write the contour for the 7-particle N$^2$MHV($=\overline{\text{NMHV}}$) amplitude by parity-conjugating the result for $k=3$. We will see that this can be viewed as having been obtained from the 6-particle N$^2$MHV($=\overline{\text{MHV}}$) amplitude by acting with an operator which adds a particle while preserving $k$, similar to the operator discussed above to derive the NMHV contour. This operator naturally generalizes to higher-$n$, and through its repeated application to the 6-particle amplitude, we obtain a closed-form result for all $n$.

As discussed in section \ref{particle_interpretation_section}, parity acts in the Grassmannian by exchanging $C$ with its dual $\widetilde{C}$, and trading all minors for their complements (see near (\ref{minor_complement})). For example, in going from $G(3,7)\to G(4,7)$, the minor $(1\,2\,3)\mapsto\overline{(1\,2\,3)}=(4\,5\,6\,7)$. Knowing this, we can immediately write down the 7-point N$^2$MHV amplitude from the NMHV amplitude given above. It is,
\begin{equation}\label{direct7nnmhv}
\hspace{0.5cm}A^{(4)}_7=\!\!\!\int\limits_{\tilde{\vf_7}=0}\!\!\frac{(3\,\,4\,\,5\,\,7)(4\,\,6\,\,7\,\,1)}{(2)(4)(6)\,\,\Big\{\underbrace{\left[(7)(3\,\,4\,\,5\,\,7)(5)\right]}_{\widetilde{f_6}}\underbrace{\left[(1)(3)(4\,\,6\,\,7\,\,1)\right]}_{\widetilde{f_7}}\Big\}},
\end{equation}
where we have used $\widetilde{f_j}$ to denote the parity-conjugates of `$f_j$', and we have used a single label in parentheses to denote any {\it consecutive} minors of $G(4,n)$---e.g., \mbox{$(2)\equiv(2\,3\,4\,5)$}. Although equation (\ref{direct7nnmhv}) is correct as written, we will find it useful to exploit the cyclic-symmetry of the Grassmannian to bring (\ref{direct7nnmhv}) into a form more reminiscent of our result for NMHV. Specifically, by rotating all particle labels in (\ref{direct7nnmhv}) by $j\mapsto j-3$, we obtain an expression remarkably similar to our form of the NMHV amplitude:

\be\label{better7nnmhv}
A^{(4)}_7=\!\!\!\int\limits_{\vf_7^{(4)}=0}\!\!\frac{(4\,7\,1\,2)(1\,3\,4\,5)}{(6)(1)(3)}\frac{1}{\underset{4567\,\,123}{\mathscr{F}}},
\ee
where we have grouped the (cyclically-rotated) parity-conjugates of $f_6$ and $f_7$ into the object 
\be\label{f4-7}
\vf_7^{(4)}\equiv\Big\{f_{7_a}^{(4)},f_{7_b}^{(4)}\Big\}\equiv\Big\{(4)(4\,7\,1\,2)(2),(5)(7)(1\,3\,4\,5)\Big\},
\ee
and where $\underset{4567\,\,123}{\phantom{{}^{}}\mathscr{F}}\equiv f_{7_a}^{(4)}\cdot f_{7_b}^{(4)}.$
%
To motivate this notation, observe that adding a particle to an $n$-point amplitude while preserving $k$ necessarily introduces $(k-2)$ new integration variables that must be fixed by the contour, and each $f^{(4)}_n$ accounts for one of these new variables. For $k=4$, therefore, it is the {\it pair} of maps $\left\{f_{7_a}^{(4)},f_{7_b}^{(4)}\right\}\equiv\vf_7^{(4)}$---taken together---which fixes the contour, and $\underset{4567\,\,123}{\phantom{{}^{}}\mathscr{F}}= f_{7_a}^{(4)}\cdot f_{7_b}^{(4)}$ which appears in the integrand. (The indices `{\footnotesize$ 4567\,\,123$}' below $\mathscr{F}$ are meant to make explicit the fact that $\mathscr{F}$ involves the {\it seven} particles numbering {\footnotesize$ 4567\,\,123$}---presented in this order. This notation will be useful below, when we consider adding particles to a general $n$-point amplitude.)

Let us now re-write the 7-particle amplitude in such a way that makes manifest that it could have been obtained by acting on the 6-particle N$^2$MHV amplitude with an `inverse-soft' operator similar to that discussed above for NMHV. Knowing $A_{7}^{(4)}$ from above, this is very easy to do:
\be
A_{7}^{(4)}=\int A_{6}^{(4)} \times\!\!\!\!\!\underset{6\to7}{\phantom{{\,}^{(4)}}\mathcal{S}^{(4)}}=\hspace{-0.3cm}\int\limits_{\vf_7^{(4)}=0}\!\!\hspace{-0.15cm}\frac{1}{(1)(2)(3)(4\,5\,6\,1)(5\,6\,1\,2)(6\,1\,2\,3)}\!\!\!\underset{6\to7}{\phantom{{\,}^{(4)}}\mathcal{S}^{(4)}},
\ee
where
\be\label{6to7}
\hspace{-0.5cm}\underset{6\to7}{\phantom{{\,}^{(k=4)}}\mathcal{S}^{(k=4)}}=\frac{(4\,5\,6\,1)(5\,6\,1\,2)(6\,1\,2\,3)(4\,7\,1\,2)(4\,2\,3\,5)(1\,3\,4\,5)}{(6\,7\,1\,2)}\frac{1}{\underset{4567\,\,123}{\phantom{{}^{}}\mathscr{F}}}.
\ee
Two important aspects of $\!\underset{6\to7}{\phantom{{^{(4)}}}\mathcal{S}^{(4)}}$ will allow it to be generalized to higher $n$ in a way which does not alter its form. First, it correctly maps the measure of $\mathcal{L}_{6,4}$ to that of $\mathcal{L}_{7,4}$: by `removing' the three minors of $G(4,6)$ which are not consecutive in $G(4,7)$---namely, $(4\,5\,6\,1), (5\,6\,1\,2)$, and $(6\,1\,2\,3)$---by including them in the numerator of $\mathcal{S}^{(4)}$; also, by adding to the measure each of the four consecutive minors of $G(4,7)$ which were not present in $\mathcal{L}_{6,4}$. One of these minors---$(6\,7\,1\,2)$---is manifest in (\ref{6to7}), while the other three minors involving particle $7$ are part of $\mathscr{F}$. Notice that {\it all} the non-consecutive minors appearing in $\mathscr{F}$ are manifestly part of the numerator of (\ref{6to7}). The second important aspect of $\mathcal{S}$ is that, by including $\mathscr{F}$ in its definition, it describes the contour of integration for the new integration variables added when going from $\mathcal{L}_{6,4}$ to $\mathcal{L}_{7,4}$ (of course, there were no integration variables for the $6$-point \mbox{N$^2$MHV($=\overline{\text{MHV}}$)} amplitude).

Let us now see how we can generalize $\!\!\!\underset{6\to 7}{\phantom{{\!\!}^(4)}\mathcal{S}^{(4)}}$ to one which adds particle $8$ to the $7$-particle amplitude. It turns out there is a very natural way of doing this. Notice that for $k=4$, the four consecutive minors of $G(4,n)$ involving $n$---which were not present in $G(4,n\smallminus1)$ and---which must be added to the measure by $\mathcal{S}$ involves exactly seven columns: $n\smallminus3,\ldots,n,1,2,3$. And because $\!\!\!\!\!\underset{6\to7}{\phantom{{}^{(4)}}\mathcal{S}^{(4)}}$ and $\!\!\underset{4567\,\,123}{\phantom{{}^{}}\mathscr{F}}$ both involve only seven fixed columns of the Grassmannian, there is a {\it canonical} way to generalize these to higher $n$. Concretely, in going from the $(n-1)$-point amplitude to the $n$-point amplitude, the inverse-soft operator must involve the minors
\be\label{newminors_k_eq4}
(n\smallminus3\,\,n\smallminus2\,\,n\smallminus1\,\,n),\qquad(n\smallminus2\,\,n\smallminus1\,\,n\,\,1),\qquad(n\smallminus1\,\,n\,\,1\,\,2),\quad\mathrm{and}\quad(n\,\,1\,\,2\,\,3)
\ee
in the denominator. It is easy to see how these can be kept manifest in $\mathscr{F}$ through its natural generalization to $\mathscr{F}_n$ by
\be\label{f4ns}
\hspace{-0.5cm}\mathscr{F}_n\!\!\equiv\!\!\underset{(n\smallminus3)(n\smallminus2)(n\smallminus1)n\,\,\,123}{\mathscr{F}}\equiv f_{n_a}^{(4)}\cdot f_{n_b}^{(4)}\ee
where 
\be\begin{split}
&f_{n_a}^{(4)}\equiv (n\smallminus3\,\,n\smallminus2\,\,n\smallminus1\,\,n)(n\smallminus3\,\,n\,\,1\,\,2)(n\smallminus3\,\,2\,\,3\,\,n\smallminus2);\\
\mathrm{and}\qquad &f_{n_b}^{(4)}\equiv(1\,\,n\smallminus2\,\,n\smallminus1\,\,n)(1\,\,n\,\,2\,\,3)(1\,\,3\,\,n\smallminus3\,\,n\smallminus2).
\end{split}
\ee
%
Notice that (\ref{f4ns}) is simply the same as (\ref{f4-7}) with the substitution $\{4,5,6,7\}\mapsto\{n\smallminus3,n\smallminus2,n\smallminus1,n\}$ while keeping $\left\{1,2,3\right\}$ fixed.

In a similar manner, we can generalize the inverse-soft operator to
\be
\hspace{-0.15cm}\underset{(n\smallminus1)\to n}{\!\!\!\!\phantom{{}^{(4)}}\mathcal{S}^{(4)}}\!\!\!\!\!=\frac{(n\smallminus3\,n\smallminus2\,n\smallminus1\,1)(n\smallminus2\,n\smallminus1\,1\,2)(n\smallminus1\,1\,2\,3)(n\smallminus3\,n\,1\,2)(n\smallminus3\,\,2\,3\,n\smallminus2)(1\,3\,n\smallminus3\,n\smallminus2)}{(n\smallminus1\,n\,1\,2)\cdot\mathscr{F}_n}\,\,.\,\,\,
\ee
By repeatedly applying this inverse-soft operator to the 6-particle N$^2$MHV amplitude, we can obtain any higher-point amplitude we like. Indeed, it is not difficult to obtain the general result for any number of particles. Doing this explicitly, we find that the $n$-particle N$^2$MHV amplitude is given by
\be\label{generaln2mhv}
\hspace{-0.cm}A_n^{(4)}=\hspace{-0.2cm}\int\limits_{\vf_n^{(4)}=0}\hspace{-0.25cm}\frac{\prod_{j=7}^{n-1}\big[\left(1\,2\,3\,j\right)\left(2\,3\,j\smallminus2\,j\smallminus1\right)\left(1\,j\smallminus2\,j\smallminus1\,j\right)\big]\prod_{j=4}^{n-3}\big[\left(1\,3\,j\,j\smallplus1\right)\left(1\,2\,j\,j\smallplus3\right)\big]}{(n\smallminus1)(1)(3)\quad\mathscr{F}_7\!\cdot \! \mathscr{F}_8\cdots \mathscr{F}_n}.
\ee
%
As we will see below, this ansatz correctly gives the 8-particle N$^2$MHV amplitude, and it does so in a remarkably-novel way---involving only four one-loop leading singularities together with sixteen two-loop (all the residues of $G(4,8)$ are at most two-loop leading singularities, \cite{ABCT:2010a}).

\subsection{The Geometry of Residues in the Grassmannian}\label{residue_geometry}

The 8-particle N$^2$MHV amplitude not only offers us an extremely good test of the ansatz (\ref{generaln2mhv}), but it also allows us the opportunity to discuss some of the more general structures involved in amplitudes (and their contours) for $k>3$. Most of these arise as a simple consequence of the fact that for $k>3$, minors of the Grassmannian are typically irreducible polynomials of degree greater than one and therefore vanish along cycles in $G(k,n)$ which multiply intersect each other (and themselves). This is true of the cycles defined by the vanishing of the (mostly non-consecutive) minors which define the tree contour in (\ref{generaln2mhv}), and it is true for the purely consecutive minors which are relevant to $\mathcal{L}_{n,k}$.

One obvious consequence of the fact that any given set of cycles can multiply-intersect is that more data is necessary to identify any particular residue than just which minors vanish on its support. And it is not true in general that distinct residues supported along the vanishing of the same set of minors are at all related. This fact becomes increasingly apparent as $n$ grows large, but is already striking for $n=9$: for example, while two of the five residues supported along by the vanishing of the minors ``$(1)(2)(3)(4)(6)(8)$'' are the leading singularities of four-mass boxes, the other three residues associated with the vanishing of these minors are simply rational functions. 

As discussed in \cite{ArkaniHamed:2009dn}, the number of isolated solutions to setting a given set of minors to zero is described by Littlewood-Richardson formula. For $k=4$ these are simply the Catalan numbers: there are generally 2 solutions to setting 4 minors to zero in $G(4,8)$; 5 solutions to setting 6 minors to zero in $G(4,9)$; 14 solutions for $G(4,10)$; 42 for $G(4,11)$; 132 for $G(4,12)$; and simple residues cease to exist for $n>12$. While we may may able to get away with labeling the 2 solutions for each set of four minors of $G(4,8)$ by simply `1' and `2,' it is clear that something more is needed in general.

As we will see below, one very powerful way to identify {\it all} the distinct residues in $G(k,n)$ is simply through the projective geometry of the Grassmannian viewed in the particle interpretation. And, perhaps even more importantly, this geometric data is closely-related to physically-important information, such as soft-limits (see \cite{ABCCK:2010}). Of course, when each column of the $C_{\alpha\,a}$-matrix is viewed as a point in $\mathbb{CP}^{k-1}$, {\it every} minor represents {\it some} geometric test. Consider the following concrete example, which arises frequently in $G(4,n)$. It is easy to show that
\be\label{egfactorization}
\hspace{-0.25cm}(2\,3\,4\,5)=(3\,4\,5\,6)=0\quad\Longrightarrow\quad\left\{\!\!\!\begin{array}{rl}\mathbf{A}&\text{all the points $\left\{2,3,4,5,6\right\}$ are coplanar;}\\
\mathbf{B}& \text{the points $\left\{3,4,5\right\}$ are collinear}.\end{array}\right.\quad
\ee
In case $\mathbf{A}$, we know as a consequence that $(2\,3\,4\,6)=0$, for example (similarly for any other choice of 4 from among $\left\{2,3,4,5,6\right\}$); and in the case of $\mathbf{B}$, we know as a consequence that $(3\,4\,5\,8)=0$ (or, more generally, $(3\,4\,5\,m)=0$ for any $m$). Notice that the natural way to test either case would be through the vanishing of a {\it non-consecutive} minor. Indeed, one way to uniquely identify every residue of the Grassmannian is to give an exhaustive list of all the minors---both consecutive and non-consecutive---which vanish on its support. (This is actually quite obvious: any point in the Grassmannian can be identified by its Pl\"ucker coordinates, which in turn can be written as a sequence of (typically non-consecutive) minors.)

One of the most remarkable features of the form of the tree-contour derived in (\ref{generaln2mhv}) is that the non-consecutive minors used to define the contour appear to {\it automatically} collapse any possible ambiguity about which particular residues are included in the contour. This turns out to be possible because for $n>7$, at least one factor among the $\mathscr{F}_n$'s given in (\ref{f4ns}) is always composed entirely of non-consecutive minors!

Another remarkable feature of the contour given in (\ref{generaln2mhv}) is that it is given {\it entirely} in terms of `simple' residues---that is, simple residues involving both consecutive and non-consecutive minors. As we will see, the $8$-point contour fixed by the contour in (\ref{generaln2mhv}) turns out to contain $9$ residues which are `composite' in terms of consecutive minors---and yet all of them arise as the {\it simple} residues of the contour. Moreover, for higher $n$, there are always $\dim(\tau)$ maps among the $\mathscr{F}$'s which define the contour, and so: {\it all residues---composites and non-composites alike---are generated as} simple residues {\it involving both consecutive and non-consecutive minors!}

\subsubsection{On the Naming of Residues}\label{naming_residues}
Before we calculate the actual residues of $G(4,8)$ which contribute to the contour given above, it is necessary for us to develop some notation to describe the residues concretely. From our discussion above, it is clear that any residue can be uniquely identified by giving a sufficiently-exhaustive list of the minors which vanish at its support. Naturally, we would like to represent this data as concisely as possible. While we will not prove it here, \mbox{(see \cite{ABCT:2010a})}, it turns out that there is a natural, physically-motivated, concise way to represent all the necessary information: {\it any} residue of $G(4,n)$ can be uniquely identified by the following:\footnote{This is only strictly true if we consider each conjugate-pair of residues associated with the leading singularities of a four-mass box as equivalent.}
\begin{enumerate}
\item a list of the consecutive minors which vanish on its support, which we write in the form, e.g., ``$(2)(4)(6)(8)$'' (where the order of these labels determines the sign of the residue);
\item all triples of consecutive, collinear points, which we indicate by a blue subscript labeling the middle of the consecutive triple; so, e.g., by ``$(2)(3)(7)(8)\soft{}{1\,4}\,$'' we mean the particular solution to $(2)(3)(7)(8)$ for which the triples $(8{\color{downstairs}1}2)$ and $(3{\color{downstairs}4}5)$ are collinear;
\end{enumerate}
and, although not strictly necessary to identify each residue, we find it useful\footnote{This is particularly relevant for $n=8$, as it is the `parity-conjugate of three points being collinear'; for higher $n$, this geometric constraint becomes increasingly constraining.} to further indicate
\begin{enumerate} \setcounter{enumi}{2}
\item all triples of consecutive points {\it whose parity conjugates} are coplanar, indicated with a red superscript labeling the middle of triple of points;  so, e.g., by ``$(2)(3)(7)(8)\soft{5\,8}{}\,$'' we mean the particular solution to $(2)(3)(7)(8)$ for which all the particles {\it in the complements of} $(4{\color{upstairs}5}6)$ {\it and} $(7{\color{upstairs}8}1)$ are coplanar---i.e.\ , for which $(78123)$ and $(23456)$ are coplanar.
\end{enumerate}

With this notation, our example (\ref{egfactorization}) can be rewritten:
\be\label{better_egfactorizations}
(2)(3)\quad\Longrightarrow\left\{\begin{array}{r}(2)(3)\soft{8}{}\\(2)(3)\soft{}{4}\end{array}\right..
\ee
As a statement about functions, (\ref{better_egfactorizations}) reads $(3)=(3)'(2)+(3)\soft{8}{}\cdot(3)\soft{}{4}\,$, which is to say, the minor $(3)$ factorizes on the support of $(2)$ (and vice versa).

It is worth keeping in mind that the collinearity and coplanarity operators are actually {\it stronger} constraints than minors alone. Specifically,
\begin{itemize}
\item each $(\cdots)\soft{}{$m$}$ implies that $(m\smallminus1\,\,{\color{downstairs}m}\,\,m\smallplus1\,\,p)=0$ for {\it any} $p$; and in particular, it implies that the minors $(m\smallminus1)=(m\smallminus2)=0$;
\item each $(\cdots)\soft{$q$}{}$ implies that any minor forming a subset of $\overline{(q\smallminus1\,\,{\color{upstairs}q}\,\,q+1)}$ vanishes; in particular, it implies that $(q\smallplus2)=\ldots=(q\smallplus n\smallminus5)=0$.
\end{itemize}

Notice that it is possible for a residue to be supported where {\it both} factors of a given minor vanish simultaneously. For example, if $(2)=0$ and both $(3)\soft{8}{}=(3)\soft{}{4}=0$, then a total of {\it three} constraints would be imposed by these two minors. Because of the symmetry between $(3)$ factorizing on $(2)$ and $(2)$ factorizing on $(3)$, we choose to indicate this extra constraint by writing $[(2)(3)]\soft{8}{4}$\,. Notice that either of the labels $()\soft{8}{}$ and $()\soft{}{4}$ imply that minors $(2)$ {\it and} $(3)$ vanish. An example of this type of composite for $n=8$ is the residue $[(2)(3)](8)\soft{8}{4}$---which will in fact contribute to the tree contour as we will see below. Similarly, if we were to know that all of the points $3, 4, 5,$ and $6$ were collinear, then we would have a residue adorned by both $()\soft{}{4}\,$ and $()
\soft{}{5}\,$; but $()\soft{}{5}$ implies that $(3)=(4)=0$, while $()\soft{}{4}$ implies $(2)=(3)=0$, and so minor $(3)$ is doubly-constrained. In this case, we would name the residue $(2)(3)^2(4)\soft{8\,1}{4\,5}\,$ (here, the coplanarity labels are a consequence of the collinearity).

Although we will not have room to discuss this here (see \cite{ABCT:2010a}), in addition to fully-specifying each distinct residue the Grassmannian, these labels also have an important, physically-motivated interpretation. They indicate how each particular residue---when viewed as a function of the kinematical variables---can be constructed out of an analogous lower-point residue in a canonical way through the action of an `inverse-soft operator' analogous to the one discussed above, but applicable to each individual residue alone and without reference to the entire amplitude. Specifically, whenever a residue involves three points being collinear in $G(k,n)$, it is canonically-related to a residue in $G(k,n-1)$ where the middle particle has been removed. Similarly, because the coplanarity of $(n-3)$ points is the parity-conjugate of three points being collinear, a coplanarity label indicates that a residue is canonically-related to a residue of $G(k-1,n-1)$ in which the labelled particle has been removed.

\subsection{The $8$-Particle N$^2$MHV Amplitude}\label{8pointsection}
We now are fully prepared to write down and compute the $8$-point N$^2$MHV amplitude as given by the general formula (\ref{generaln2mhv}). Explicitly, we have
\be\label{8ptcontour}
\hspace{-0cm}A_8^{(4)}=\!\!\!\!\!\int\limits_{\vf_8^{(4)}=0}\!\!\!\!\frac{(5\,6\,7\,1)(7\,1\,2\,3)(2\,3\,5\,6)(1\,2\,4\,7)(1\,3\,4\,5)(1\,2\,5\,8)(1\,3\,5\,6)}{(7)(1)(3)\,\,\mathscr{F}_7\cdot\mathscr{F}_8},
\ee
where, from (\ref{f4ns}),
\be\label{contourfs}
\begin{split}
\mathscr{F}_7&=\Big[(4)(4\,7\,1\,2)(2)\Big]\times\Big[(1\,2\,3\,7)(3\,4\,5\,1)(5\,6\,7\,1)\Big],\\
\hspace{-1.5cm}\mathrm{and}\qquad \mathscr{F}_8&=\Big[(5)(5\,8\,1\,2)(5\,2\,3\,6)\Big]\times\Big[(6)(8)(1\,3\,5\,6)\Big].
\end{split}
\ee
This multidimensional contour integral involves a few subtleties beyond those already encountered for NMHV contours. As discussed at length above, the principle new subtlety encountered for $k=4$ is that the minors which define the contour are generically quadratic polynomials, whose cycles of zeros typically intersect each other (and themselves) multiply. Another novelty first encountered for $k=4$ is that it is possible for some of the minors within the $f_i$'s to factorize on a solution of the others, leading to multiple branches which can sometimes can have very different structures. These potential subtleties are best understood through example. Therefore, in the next subsection, we will work through a number of the contributions (and potential contributions) to the tree amplitude coming from the contour above, trying to sample all of the possible types of contributions. 

Before we begin our series of examples, it is useful to lay-out the form we expect the answer to take, and the type of calculation that will be involved in the evaluating (\ref{8ptcontour}). Because setting any 4 minors of $G(4,8)$ to zero will typically have 2 isolated solutions, we may first expect that by pairing any of the three minor-factors of the $f_i$'s together, we would find $\lesssim3^4*2=162$ isolated poles in the Grassmannian `encompassed' by the contour. Of course, the numerator of (\ref{8ptcontour}) ensures that any pole generated by the $f_i$'s which is not a pole of {\it consecutive minors} will have a vanishing residue. Therefore, we expect that the vast-majority of isolated solutions to $f_i=0,$ for $i=1,\ldots,4$ will not contribute anything to the amplitude. Indeed, it turns out that among all the $3^4$ choices of factors from among the $f_i$'s (and all of their multiple solutions), only 20 poles will contribute a non-vanishing residue to the contour---and these terms have been checked to add-up to precisely the 8-particle amplitude, matching right-down to the sign of every term.

\subsubsection{Example Contributions from the Contour}
In order to gain some understanding of how each of the 20 non-vanishing residues are generated by the contour, it is worthwhile to analyze a few examples in detail. Let us start by rewriting the maps $f_i$ which define the contour in a slightly more transparent way:
\be
\begin{split}
&f_1=\left[(2\,3\,4\,5)(4\,5\,6\,7)(7\,1\,2\,4)\right],\qquad &&f_3=\left[(5\,6\,7\,8)(2\,3\,5\,6)(8\,1\,2\,5)\right],\\
&f_2=\left[(1\,2\,3\,7)(3\,4\,5\,1)(5\,6\,7\,1)\right],\quad\quad&&f_4=\left[(6\,7\,8\,1)(8\,1\,2\,3)(3\,5\,6\,1)\right].
\end{split}\quad
\ee
Notice that the contour is naturally composed some $3^4$ parts coming from the simultaneous vanishing of any choice of factors from among the $f_i$'s. However, because $f_2$ is entirely composed of non-consecutive minors, most poles of the contour will have vanishing residue and contribute nothing to the tree amplitude. The exceptional cases are those for which the solution to $f_1=\ldots=f_4=0$ is also a pole in $\mathcal{L}_{8,4}$. The complete list of such contributions is given in Table \ref{8-point-residue-table} at the end of this section. Each of these contributions is quite easy to understand geometrically, and considering a few exercises in particular will illustrate the role of projective geometry in the general contour.

\vspace{0.3cm}
\begin{itemize}
\item[\mbox{\textbullet\hspace{1.25cm}}] \hspace{-1.25cm} $(2\,3\,4\,5)(3\,4\,5\,1)(2\,3\,5\,6)(1\,3\,5\,6)\Longrightarrow (2)(3)^2(4)\soft{8\,1}{4\,5}$
\end{itemize}
Notice that this choice of minors from the $f_i$'s includes only one consecutive minor, $(2\,3\,4\,5)$, together with the three non-consecutive minors $(1\,3\,4\,5), (2\,3\,5\,6),$ and $(1\,3\,5\,6)$. The important thing to notice about these four minors is that they all involve points $3$ and $5$. This means that the geometry problem at hand is merely the classic problem of Schubert calculus of finding the set of lines---in this case the lines $`[3\,5]$'---which intersect four given lines in $\mathbb{P}^3$.

Here, the four lines which $[3\,5]$ must intersect are $[1\,4],[4\,2],[2\,6],[6\,1]$. Notice that these four lines mutually intersect at points $4,2,6,$ and $1$, forming a closed loop. This is illustrated on the left-hand side of the figure below. It is not hard to see that the only two solutions are those shown on the right-hand side of the same figure, $[3\,5]_{\rm{A}}$ and $[3\,5]_{\rm{B}}$.

The solution $[3\,5]_{\rm{A}}$ involves all four points $\left\{1,2,3,5\right\}$ being collinear. While this configuration implies that minors $(8)$ and $(1)$ vanish, it does not provide a fourth constraint coming from a consecutive minor, and therefore the residue associated with this pole will vanish in the contour.

\begin{figure}[t]
\centering\includegraphics[scale=1]{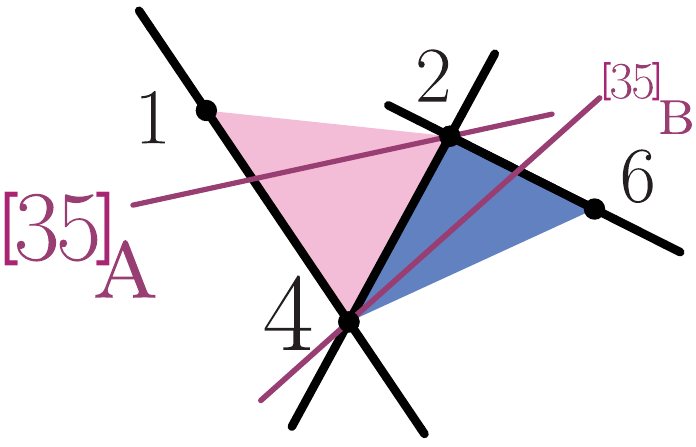}\caption{The two classes of solutions to setting minors $(2\,3\,4\,5), (1\,3\,4\,5),$  and $(2\,3\,5\,6)$, to zero. In solution {\bf A}, line $[3\,5]$ lies on the plane $[1\,2\,4]$ and passes through the point $2$; for {\bf B}, the line $[3\,5]$ lies on the plane $[6\,2\,4]$ and passes through the point $4$.\label{fig2}}\end{figure}

The solution $[3\,5]_{\rm{B}}$, on the other hand, involves all the points $\left\{3,4,5,6\right\}$ being collinear. Recall that when $3,4,5$ are collinear, minors $(2)$ and $(3)$ vanish, and when $4,5,6$ are collinear, minors $(3)$ and $(4)$ vanish. Thus, the minor $(3)$ is doubly-constrained, and we find that this geometric configuration contributes the residue $(2)(3)^2(4)\soft{8\,1}{4\,5}$ to the amplitude.
\[
\includegraphics[scale=0.85]{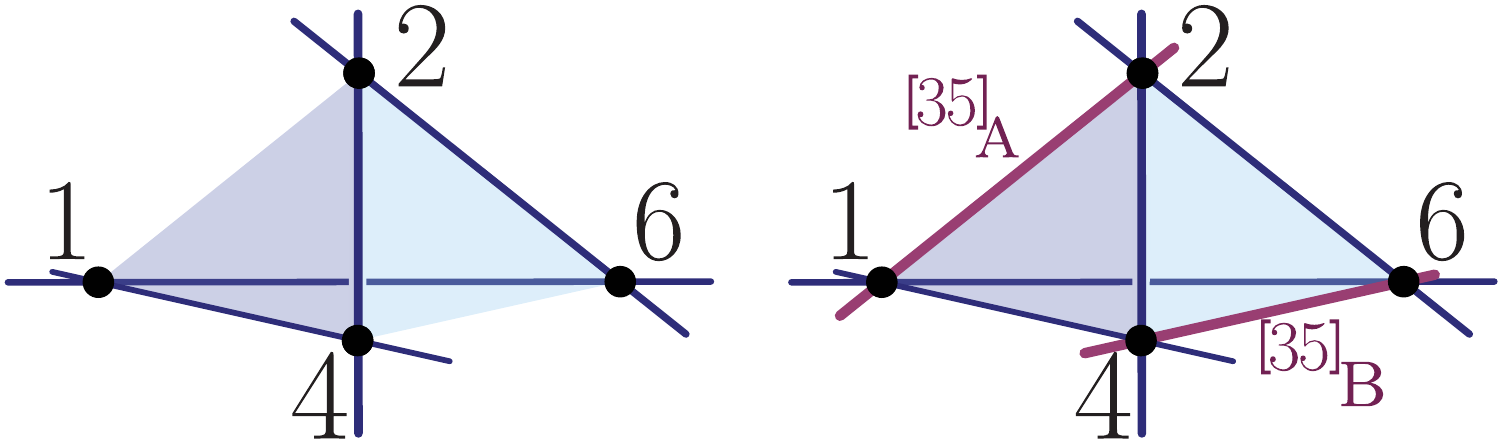}\nonumber
\]
\newpage
\begin{itemize}
\item[\mbox{\textbullet\hspace{1.25cm}}] \hspace{-1.25cm} $(2\,3\,4\,5)(3\,4\,5\,1)(2\,3\,5\,6)(8\,1\,2\,3)\Longrightarrow [(2)(3)](8)\soft{8}{4}$
\end{itemize}
The first three minors of this problem are the same as in the last problem. Let us start by considering these minors by themselves. As before, because all three minors involve the particles $3$ and $5$, we are looking for the configurations of lines $[3\,5]$ which intersect the three given lines $[1\,4],[4\,2],$ and $[2\,6]$. There are two families of such solutions which are illustrated in Figure \ref{fig2}. Specifically, these two solutions are:
\begin{itemize}
\item[\bf{A}] the line $[3\,5]$ passes through the point $2$ and lies on the plane $[1\,4\,2]$, or
\item[\bf{B}] the line $[3\,5]$ passes through the point $4$ and lies on the plane $[6\,4\,2]$.
\end{itemize}
Now let us consider imposing the additional constraint $(8\,1\,2\,3)=0$ to each of the two cases. In case {\bf A}, $(8\,1\,2\,3)=0$ implies that the line $[8\,1]$ intersect $[2\,3]=[2\,5]=[3\,5]$. The only configuration then, is where the line $[3\,5]$ lies along $[1\,2]$, which was the same case we encountered in the previous geometry problem---and one that does not involve enough consecutive minors to contribute to the amplitude.

For case {\bf B}, the line $[8\,1]$ will intersect the plane $[2\,4\,6]$ at some point through which $[3\,5]$ must pass; this will fix the angular freedom of $[3\,5]$ on the plane $[2\,4\,6]$. Therefore, we have that $3,4,$ and $5$ are collinear, {\it and} the points $2,3,4,5,6$ are coplanar. Both of these conditions set the minors $(2)$ and $(3)$ to zero, and so the two minors $[(2)(3)]\soft{8}{4}$ contribute a total of three constraints. When combined with minor $(8)$, we obtain the composite residue $[(2)(3)](8)\soft{8}{4}\,$.

\newpage
\begin{itemize}
\item[\mbox{\textbullet\hspace{1.25cm}}] \hspace{-1.25cm}$(2\,3\,4\,5)(5\,6\,7\,1)(5\,6\,7\,8)(8\,1\,2\,3)\Longrightarrow (2)(4)(5)(8)\soft{}{6}\quad\mathrm{and}\quad(2)(6)(5)(8)\soft{3}{}$
\end{itemize}
Recall how consecutive minors factorized in the example (\ref{better_egfactorizations}). Just as in that case, because minors $(5\,6\,7\,8)$ and $(5\,6\,7\,1)$ overlap on three columns, we may conclude that, on the support of $(5)$, $(5\,6\,7\,1)\to(6)\soft{3}{}\cdot(4)\soft{}{6}\,$. What this means for this case is that the two solutions to $(5\,6\,7\,1)=(5\,6\,7\,8)=0$ are $(4)(5)\soft{}{6}\,$ and $(5)(6)\soft{3}{}\,$. Combining these two constraints with the minors $(2)$ and $(8)$ from $f_1$ and $f_4$, respectively, we find that the two solutions are: $(2)\Big[(4)\soft{}{6}+(6)\soft{3}{}\,\Big](5)(8)=(2)(4)(5)(8)\soft{}{6}+(2)(6)(5)(8)\soft{3}{}\,$.

Before we move on to the next example, it is worth emphasizing that the {\it ordering} of minors appearing in the residue ``$(2)(6)(5)(8)\soft{3}{}\,$'' was fixed by the ordering of the $f_i$'s: minor $(5\,6\,7\,1)$ appearing in $f_2$ contributed the `$(6)$,' while $f_3$ contributed minor $(5)$. This completely fixes the signs of the tree-contour.

\begin{figure}[b]
\centering\includegraphics[scale=0.85]{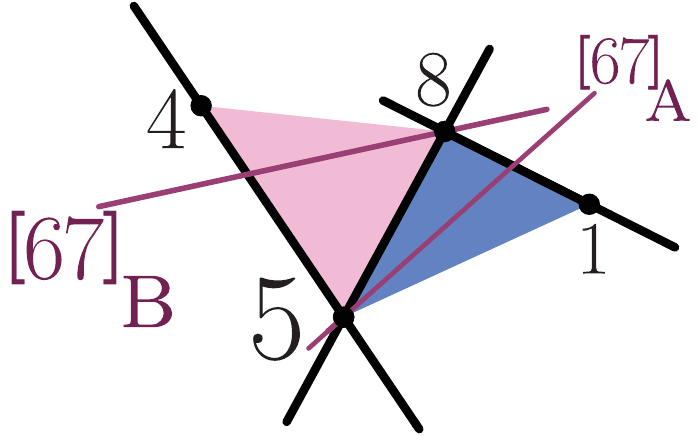}\caption{The two classes of solutions to setting minors $(4\,5\,6\,7)=(5\,6\,7\,8)=(6\,7\,8\,1)=0$, where the possible configurations for the line $[6\,7]$ are indicated.\label{fig3}}
\end{figure}

\begin{itemize}
\item[\mbox{\textbullet\hspace{1.25cm}}] \hspace{-1.25cm} $(4\,5\,6\,7)(5\,6\,7\,1)(5\,6\,7\,8)(6\,7\,8\,1)\Longrightarrow (4)(5)^2(6)\soft{2\,3}{6\,7}$
\end{itemize}
Let us start this problem by first considering the three minors $(4\,5\,6\,7)$, $(5\,6\,7\,8)$ and $(6\,7\,8\,1)$. Here, we have that the line $[6\,7]$ must intersect the three lines $[4\,5],[5\,8],$ and $[8\,1]$. This case should be familiar from before, and is illustrated in Figure \ref{fig3}. There are two infinite families of solutions:
\begin{itemize}
\item[\bf{A.}] the line $[6\,7]$ passes through the point $5$ and lies on the plane $[1\,5\,8]$, or
\item[\bf{B.}] the line $[6\,7]$ passes through the point $8$ and lies on the plane $[4\,5\,8]$.
\end{itemize}
Let us first consider case {\bf A}. Here, we see that there is an apparent problem: when the points $\left\{5,6,7,8,1\right\}$ are coplanar, we {\it automatically} have that minor $(5\,6\,7\,1)=0$, and so $f_2$ vanishes everywhere over this entire infinite `sheet' which solves the first three constraints! Clearly, when $f_2=0$ everywhere over a surface, it does not generate a transversally-supported pole. Said another way, $f_2$ vanishes {\it trivially} for this class of solutions, and only because the non-consecutive minor \mbox{$(5\,6\,7\,1)$} vanishes. But this also vanishes everywhere in the numerator and so it effectively imposes no constraint at all.

In case {\bf B}, however, $(5\,6\,7\,1)$ is {\it not} manifestly zero. Here, in fact, the vanishing of $(5\,6\,7\,1)$ imposes the non-trivial constraint that $[6\,7]$ intersects the point 5. Notice that this is actually where {\it both} of the factors of minor $(5\,6\,7\,1)=0$---one factor which tests the coplanarity of the points $\{5,6,7,1\}$ and the other which tests the collinearity of the points $\{5,6,7\}$. For this solution, the line $[6\,7]$ must lie along the line $[5\,8]$, and hence the points $\left\{5,6,7,8\right\}$  are all collinear! Similar to our first example above, the collinearity of $\left\{5,6,7\right\}$ implies that minors $(4)$ and $(5)$ vanish, while the collinearity of $\{6,7,8\}$ implies that the minors $(5)$ and $(6)$ vanish. This leads to the composite residue $(4)(5)^2(6)\soft{2\,3}{6\,7}\,$.

\begin{table}[h]
\centering\hspace{0.cm}\begin{tabular}{|l|l|}
\hline\,\,\,\,Residue&\hspace{0.55cm}Geometry Problem:\\
&$\hspace{0.475cm} f_1\hspace{0.75cm}f_2\hspace{0.75cm}f_3\hspace{0.75cm}f_4$\\
\hline
$\rule[-0.ex]{0pt}{3ex}(2)(3)^2(4)\soft{8\,1}{4\,5}$&$(2345)(3451)(2356)(1356)$\\
$\rule{0pt}{3ex}[(2)(3)](6)\soft{8}{4}$&$(2345)(3451)(2356)(6781)$\\
$\rule{0pt}{3ex}[(2)(3)](8)\soft{8}{4}$&$(2345)(3451)(2356)(8123)$\\
$\rule{0pt}{3ex}(2)[(5)(6)]\soft{3}{7}$&$(2345)(5671)(5678)(6781)$\\
$\rule{0pt}{3ex}(2)[(7)(8)]\soft{5}{1}$&$(2345)(1237)(8125)(8123)$\\
$\rule{0pt}{3ex}(4)(5)^2(6)\soft{2\,3}{6\,7}$&$(4567)(5671)(5678)(6781)$\\
$\rule{0pt}{3ex}[(4)(5)](8)\soft{2}{6}$&$(4567)(5671)(5678)(8123)$\\
$\rule{0pt}{3ex}(4)[(7)(8)]\soft{5}{1}$&$(4567)(1237)(8125)(8123)$\\
$\rule{0pt}{3ex}(6)(7)^2(8)\soft{4\,5}{8\,1}$&$(7124)(7123)(8125)(8124)$\\
$\rule{0pt}{3ex}(2)(1)(5)(8)\soft{7}{2}$&$(2345)(7123)(5678)(8123)$\\[0.5ex]\hline
\end{tabular}\hspace{0.754cm}\begin{tabular}{|l|l|}\hline\,\,\,\,Residue&\hspace{0.55cm}Geometry Problem:\\
&$\hspace{0.475cm} f_1\hspace{0.75cm}f_2\hspace{0.75cm}f_3\hspace{0.75cm}f_4$\\
\hline
$\rule{0pt}{3ex}(2)(3)(5)(6)\soft{}{4\,7}$&$(2345)(3451)(5678)(6781)$\\
$\rule{0pt}{3ex}(2)(5)(3)(6)\soft{3\,8}{}$&$(2345)(5671)(2356)(6781)$\\
$\rule{0pt}{3ex}(2)(3)(5)(8)\soft{}{4}$&$(2345)(3451)(5678)(8123)$\\
$\rule{0pt}{3ex}(2)(3)(7)(8)\soft{}{1\,4}$&$(2345)(3451)(8125)(8123)$\\
$\rule{0pt}{3ex}(2)(7)(3)(8)\soft{5\,8}{}$&$(2345)(1237)(2356)(8123)$\\
$\rule{0pt}{3ex}(2)(4)(5)(8)\soft{}{6}$&$(2345)(5671)(5678)(8123)$\\
$\rule{0pt}{3ex}(2)(6)(5)(8)\soft{3}{}$&$(2345)(5671)(5678)(8123)$\\
$\rule{0pt}{3ex}(2)(7)(5)(8)\soft{5}{}$&$(2345)(1237)(5678)(8123)$\\
$\rule{0pt}{3ex}(4)(5)(7)(8)\soft{}{1\,6}$&$(4567)(5671)(8125)(8123)$\\
$\rule{0pt}{3ex}(4)(7)(5)(8)\soft{2\,5}{}$&$(4567)(1237)(5678)(8123)$\\[0.5ex]
\hline
\end{tabular}\caption{All of the non-vanishing residues contributing to the 8-point N$^2$MHV amplitude as given in (\ref{8ptcontour}), and the corresponding `geometry problem' that gives rise to each.\label{8-point-residue-table}}
\end{table}

\newpage

\subsubsection{Summary of $8$-Point N$^2$MHV Results}

Continuing to solve the various geometry-problems in this manner, we would eventually find that the complete contour given in (\ref{8ptcontour}) contributes only 20 non-vanishing residues to the tree-amplitude. These 20 terms are as follows:
\vspace{-10ex}\be\label{8n2mhvcontour}
\hspace{-0.75cm}A_{8}^{(4)}=\hspace{-0.05cm}\begin{array}{llllllllll}\rule[-0.ex]{0pt}{13ex}&\rule[-0.ex]{0pt}{3ex}\,\,(2)(3)^2(4)\softup{8\,1}\softdown{4\,5}&\!\!\!+\!\!\!&\,\,[(2)(3)](6)\soft{8}{4}&\!\!\!+\!\!\!&\,\,[(2)(3)](8)\soft{8}{4}&\!\!\!+\!\!\!&\,\,(6)(7)^2(8)\soft{4\,5}{8\,1}&\!\!\!+\!\!\!&\,\,(2)[(7)(8)]\soft{5}{1}\\[0.5ex]
\!\!\!+\!\!\!&\,\,(4)[(7)(8)]\soft{5}{1}&\!\!\!+\!\!\!&\,\,(2)[(5)(6)]\soft{3}{7\phantom{\,5}}&\!\!\!+\!\!\!&\,\,(4)(5)^2(6)\soft{3\,2}{7\,6}&\!\!\!+\!\!\!&\,\,[(4)(5)](8)\soft{2}{6}&\!\!\!+\!\!\!&(2)(1)(5)(8)\soft{7}{2}\\[0.5ex]
\!\!\!+\!\!\!&(2)(3)(5)(6)\soft{}{4\,7}&\!\!\!+\!\!\!&(2)(5)(3)(6)\soft{3\,8}{}&\!\!\!+\!\!\!&(2)(3)(5)(8)\soft{}{4}&\!\!\!+\!\!\!&(2)(3)(7)(8)\soft{}{1\,4}&\!\!\!+\!\!\!&(2)(7)(3)(8)\soft{5\,8}{}\\[0.5ex]
\!\!\!+\!\!\!&(2)(4)(5)(8)\soft{}{6\phantom{\,7}}&\!\!\!+\!\!\!&(2)(6)(5)(8)\soft{3\phantom{\,8}}{}&\!\!\!+\!\!\!&(2)(7)(5)(8)\soft{5}{}&\!\!\!+\!\!\!&(4)(5)(7)(8)\soft{}{1\,6}&\!\!\!+\!\!\!&(4)(7)(5)(8)\soft{2\,5}{}
\end{array}\hspace{0.5cm}
\ee
We have checked that this correctly matches the result calculated in field theory. The geometric origin of each of these terms is summarized in Table \ref{8-point-residue-table}.

One of the remarkable features of (\ref{8n2mhvcontour}) is that among all the residues of the contour, only 4 are primitive one-loop leading singularities---namely, $(2)(3)^2(4)\soft{8\,1}{4\,5}\,\,$, $\,(4)(5)^2(6)\soft{2\,3}{6\,7}\,\,,\,(6)(7)^2(8)\soft{4\,5}{8\,1}\,\,$, and $(2)(1)(5)(8)\soft{7}{2}\,$, of which the first three are cyclic-variants of the function `$X$' of \cite{Britto:2004ap}, while the last is cyclically-related to `$V$' (see also \cite{ArkaniHamed:2009dn}). All the other residues appearing in (\ref{8n2mhvcontour}) are two-loop leading singularities; these and similar facts will be discussed at length in a paper specifically focused on residues in $G(k,n)$ for $k\geq4$, \cite{ABCT:2010a}.

One may naturally wonder if there is any similarity between the structure of the tree-contour in (\ref{8n2mhvcontour}) and the even/odd structure of the NMHV contour. In some sense there is: knowing how each of the factors of each $f_i$ contributes to the non-vanishing terms in (\ref{8n2mhvcontour}), we find that the tree-contour can be re-written (somewhat schematically) as,
\be\label{better-8-pt-amplitude}
\hspace{-0.5cm}A_{8}^{(4)}\!\!=\Big[(2)\smallplus(4)\smallplus(6)\soft{}{8\,1}\Big]\!\!\left[(5)\soft{3}{}\smallplus(7)\soft{5}{}\smallplus(1)\soft{7}{2}\smallplus(3)\soft{}{4}\smallplus(5)\soft{}{6}\right]\!\!\left[(3)\soft{}{5}\smallplus(7)\soft{}{1}\smallplus(5)\smallplus(7)\soft{4}{}\smallplus(3)\soft{8}{}\,\right]\!\!\left[(4)\soft{8\,1}{}\smallplus\,(6)\smallplus(8)\right].\quad\nonumber
\ee
By expanding this formula and keeping only the terms that are consistent with the constraints implied by the collinearity/coplanarity operators, precisely the 20 terms of the tree-contour given in (\ref{8n2mhvcontour}) are found.

\newpage
\subsection{Connection to the Twistor String}

We can now take our proposal for all N$^2$MHV amplitudes and deform it along the lines explained in section 5 in order to get an integral over the Grassmannian localized on $C$-matrices of the Veronese form. In other words we take
\be\label{fino}
\hspace{-0.cm}A_n^{(4)}=\int\limits_{\vS_n^{(4)}=0} \hspace{-0.15cm}\frac{\mathscr{H}_{n}^{(4)}}{
\quad\mathscr{S}^{(4)}_7\!\cdot \! \mathscr{S}^{(4)}_8\cdots \mathscr{S}^{(4)}_n}\,\,,
\ee
where \be
\hspace{-0.75cm}\mathscr{H}^{(4)}_n=\frac{\displaystyle\prod_{j=7}^{n-1}\!\Big[\!\!\left(1\,2\,3\,j\right)\!\left(2\,3\,j\tm2\,\,j\tm1\right)\!\left(1\,j\tm2\,\,j\tm1\,\,j\right)\!\!\Big]\!\prod_{j=4}^{n-3}\!\Big[\!\!\left(1\,3\, j\,\, j\smallplus1\right)\!\left(1\,2\, j\,\,j\smallplus3\right)\!\left(1\,3\, j\,\, j\smallplus2\right)\!\left(1\,2\, j\,\, j\smallplus2\right)\!\!\Big]}{(n\smallminus1)\,(1)\,(3)}\,\,\,,\,\,\,\,\,\hspace{0.05cm}
\ee
and $\vS_n^{(4)}\equiv\left\{S_{7_a}^{(4)},S_{7_b}^{(4)},\ldots,S_{n_a}^{(4)},S_{n_b}^{(4)}\right\}$ with 
\be
\begin{split}
S_{k_a}^{(4)}\equiv&\phantom{\,-\,}(k\smallminus3\,\,k\smallminus2\,\,k\smallminus1\,\,k)(k\smallminus3\,\,k\,\,1\,\,2)
(k\smallminus3\,\,2\,\,3\,\,k\smallminus2)(k\smallminus3\,k\smallminus1\,1\,3)\\
&-(k\smallminus3\,k\smallminus1\,k\,1)
(k\smallminus3\,1\,2\,3)(k\smallminus3\,3\,k\smallminus2\,k\smallminus1)(k\smallminus3\,k\,2\,k\smallminus2); \\
\mathrm{and}\qquad S_{k_b}^{(4)}\equiv&\phantom{\,-\,}(1\,\,k\smallminus2\,\,k\smallminus1\,\,k)(1\,\,k\,\,2\,\,3)(1\,\,3\,\,k\smallminus3\,\,k\smallminus2)
(1\,k\smallminus1\,2\,k\smallminus3)\\
&-(1\,k\smallminus1\,k\,2)(1\,2\,3\,k\smallminus3)(1\,k\smallminus3\,k\smallminus2\,k\smallminus1)(1\,k\,3\,k\smallminus2);
\end{split}
\ee
and each $\mathscr{S}^{(4)}_k$ represents the product the two Veronese operators $S_{k_a}^{(4)}\cdot S_{k_b}^{(4)}$.

The natural question at this point is whether this form agrees with the twistor string formula.
In order to check this we take the twistor string formula equation (\ref{fofc}) and gauge fix GL$(2)$  using $\xi_1$, $\rho_1$, $\rho_2$ and $\rho_3$ and gauge fix GL$(4)$ to some link representation. Therefore we get an integral of the form \cite{Spradlin:2009qr}
\be
\label{rita}
J_{\mathrm{GL(2)}}\int \frac{d\rho_4d\rho_5\cdots d\rho_n}{(\rho_1-\rho_2)(\rho_2-\rho_3)\cdots (\rho_n-\rho_1)}\int \prod_{i=2}^n \frac{d\xi_i}{\xi_i} \prod_{i,J}\delta \left( c_{iJ} - \frac{\xi_i\xi_J}{\rho_i-\rho_J}\right)
\ee
where $J_{\mathrm{GL(2)}} = \xi_1(\rho_1-\rho_2)(\rho_2-\rho_3)(\rho_3-\rho_1)$. Here, $i$ runs over four indices (the ones chosen for the link representation), while $J$ runs over the remainder $n-4$.
And we can now expand around any fixed configuration $\hat c_{iJ} = \hat \xi_i\hat\xi_J/(\hat\rho_i-\hat\rho_J)$. In other words, we may take \mbox{$c_{iJ} = \hat c_{iJ} + h_{iJ}^a \epsilon_a$} where $h_{iJ}^a$ are some generic functions of $\hat\rho$'s and $\hat\xi$'s, where $a=1,\ldots,2(n-6)$.
Now we take the system of $4(n-4)$ equations given by the $\delta$-functions as a system that `locks' all $2(n-6)$ $\epsilon$'s to zero and all $n-3$ $\rho$'s and all $n-1$ $\xi$'s to their hatted values. This means that (\ref{rita}) becomes
\be
{\cal I}_{\rm Twistor-String} \equiv \frac{\hat\xi_1(\hat\rho_1-\hat\rho_2)(\hat\rho_2-\hat\rho_3)(\hat\rho_3-\hat\rho_1)}{(\hat\rho_1-\hat\rho_2)
(\hat\rho_2-\hat\rho_3)\cdots (\hat\rho_n-\hat\rho_1)}\times J_{4(n-4)}(\hat\rho,\hat\xi,0),
\ee
where $J_{4(n-4)}(\hat\rho,\hat\xi)$ is the Jacobian of the $4(n-4)$ equations $E_{iJ} = \hat \xi_i\hat\xi_J/(\hat\rho_i-\hat\rho_J) + h_{iJ}^a \epsilon_a - \frac{\xi_i\xi_J}{\rho_i-\rho_J}$ evaluated on the hatted values and $\epsilon  =0 $---i.e.\ ,
\be
J_{4(n-4)} = \frac{\partial (E_{iJ})}{\partial (\epsilon's, \xi's,\rho's)}.
\ee
On the Grassmannian side, we gauge-fix GL$(4)$ in the same way and expand $c_{iJ} = \hat c_{iJ} + h_{iJ}^a \epsilon_a$. Using this expansion, each of the $2(n-6)$ Veronese operators becomes linear in $\epsilon$'s to leading order. Therefore we can evaluate the integral (\ref{fino}) and obtain
\be
{\cal I}_{\rm G} \equiv \left.\mathscr{H}_n^{(4)}\right|_{c_{iJ} = \hat c_{iJ}} \times J_{2(n-6)},
\ee
where the Jacobian $J_{2(n-6)}$ is given by
\be
\left.\frac{\partial ({S}^{(4)}_{7_a},\ldots , S^{(4)}_{n_b}) }{\partial(\epsilon_1,\ldots ,\epsilon_{2(n-6)})}\right|_{\epsilon = 0}\,\,.
\ee
We have checked that ${\cal I}_{\rm Twistor-String} = {\cal I}_{\rm G}$ for $n=7,8,9$ and $10$. It would be interesting to find a general proof for all $n$.

\section{Discussion}\label{discussion_section}\setcounter{equation}{0}

The expression for ${\cal L}_{n,k}$ as a contour integral over the
Grassmannian $G(k,n)$ makes the  Yangian symmetry
\cite{Drummond:2009fd} of ${\cal N}=4$ SYM manifest. Since conformal
and dual superconformal symmetries act on mutually non-local spaces,
it is not surprising that each individual residue of ${\cal
L}_{n,k}$ does not have a good local space-time interpretation;
rather, there is by now a great deal of evidence for the conjecture
of \cite{ArkaniHamed:2009dn}, that the residues compute leading
singularities of scattering amplitudes at all loop orders. Even at
tree-level, however, a central issue is to understand how local
space-time physics emerges. As we saw in \cite{ArkaniHamed:2009sx},
for the special contours associated with the tree amplitude, a
canonical contour deformation can expose the spacetime Lagrangian in
light-cone gauge via the CSW/Risager rules. But the more fundamental
question remains: what is invariantly special about this contour? Is
there a question intrinsic to the Grassmannian that singles it out?
In this paper we have clearly seen the outlines of the answer to
this question. Demanding that our integral over $G(k,n)$ has a
``particle interpretation" {\it in the Grassmannian} picks out a
contour that gives us the tree amplitudes with a good space-time
interpretation. The notion of a particle interpretation in the
Grassmannian seems more
primitive and fundamental than locality in space-time, since it is formulated in a setting
that exhibits all the symmetries of the
theory. Unifying the residues of $\Gamma^{{\cal L}}_{n,k}$ into a
single variety leads to an ``add one at a time" particle interpretation which makes the Yangian symmetry manifest. 
The Veronese particle interpretation is equivalent to the  connected prescription for
twistor string theory. Quite beautifully, these apparently different
sorts of Grassmannian theories are simply related by a deformation
parameter $t$. The theory at $t=0$ corresponds directly to
the unified form of ${\cal L}_{n,k}$ with contour $\Gamma^{{\cal L}}_{n,k}$,
while the connected prescription amplitude ${\cal T}_{n,k}$
corresponds to $t=1$.
Thinking of $t$ as analogous to RG time, ${\cal L}_{n,k}$ is like
the ``ultraviolet" theory, where the full Yangian symmetry is
manifest, while ${\cal T}_{n,k}$ is akin to the confined description
in the infrared, where the ``macroscopic" properties of the
collection of residues---especially the cyclic symmetries and $U(1)$-decoupling identities---are manifest. For NMHV amplitudes a simple
residue theorem demonstrates $t$-independence, and we expect a
generalization of this argument should be possible for all $k$.
Indeed, while have restricted our discussion in this paper to NMHV and
N$^2$MHV amplitudes,
we fully expect the basic physical picture for tree amplitudes we have
presented in this paper to generalize for arbitrary $k$. A number of
new issues arise for $k>4$---in particular the distinction between the
more natural localization in $\mathbb{CP}^{k-1}$ versus localization
in the $\mathbb{CP}^3$ of twistor space first becomes apparent for
$k=5$---and we will return to examine these issues in future work.

We have focused exclusively on  tree amplitudes in this paper, yet
clearly the most exciting feature of ${\cal L}_{n,k}$ is that it
contains all-loop information. Can the ``particle interpretation"
picture in the Grassmannian be generalized to include full
loop-level amplitudes, not just leading singularities?

~\\
{\bf Note added}: as our manuscript was being prepared, Nandan, Volovich and
Wen published a paper studying a GL$(3)$ invariant form of the connected
prescription. They also noted that a deformation of this object leads
to $\mathcal{L}_{n,3}$, and gave a residue theorem argument  for $t$ independence.\\

\section*{Acknowledgments}

We thank Louise Dolan, Peter Goddard and Edward Witten for stimulating discussions. N.A.-H. is supported by the DOE under grant DE-FG02-91ER40654, F.C. is supported in part by the NSERC of Canada and MEDT of Ontario and by The Ambrose Monell Foundation. J.T. is supported by the U.S. Department of State through a Fulbright Award.

%

\providecommand{\href}[2]{#2}\begingroup\raggedright\endgroup

\end{document}